\documentclass[journal]{IEEEtran}
\pdfoutput = 1
\usepackage[T1]{fontenc}
\usepackage[latin9]{inputenc}
\usepackage{textcomp}
\usepackage{amsmath}
\usepackage{amsthm}
\usepackage{amssymb}
\usepackage{graphicx}
\usepackage{bm}
\usepackage{booktabs}
\usepackage{multirow}
\usepackage{rotating}
\usepackage{subfigure}
\usepackage{stfloats}
\usepackage[vlined,boxed,ruled,linesnumbered,resetcount]{algorithm2e}
\usepackage{subfigure}
\usepackage{xcolor}
\usepackage[noadjust,sort,compress]{cite}
\usepackage{setspace}
\usepackage{caption}
\usepackage{subcaption}
\usepackage{threeparttable}
\begin{document}
	\newtheorem{lemma}{Lemma}
	\newtheorem{corol}{Corollary}
	\newtheorem{theorem}{Theorem}
	\newtheorem{proposition}{Proposition}
	\newtheorem{definition}{Definition}
	\newcommand{\e}{\begin{equation}}
		\newcommand{\ee}{\end{equation}}
	\newcommand{\eqn}{\begin{eqnarray}}
		\newcommand{\eeqn}{\end{eqnarray}}

	\newenvironment{shrinkeq}[1]
	{ \bgroup
		\addtolength\abovedisplayshortskip{#1}
		\addtolength\abovedisplayskip{#1}
		\addtolength\belowdisplayshortskip{#1}
		\addtolength\belowdisplayskip{#1}}
	{\egroup\ignorespacesafterend}
	\title{Pseudo-Random TDM-MIMO FMCW Based Millimeter-Wave Sensing and Communication Integration for UAV Swarm}
	\author{Yi Tao, Zhen Gao, {\textit{Member, IEEE}}, Zhuoran Li, Ziwei Wan, Tuan Li, Chunli Zhu, Lei Chen, Guanghui Wen, {\textit{Senior Member, IEEE}}, Dezhi Zheng, {\textit{Member, IEEE}}, Dusit Niyato, {\textit{Fellow, IEEE}}
		\thanks{The work of Zhen Gao was supported in part by the Natural Science Foundation of China (NSFC) under Grants 62471036 and U2233216, Shandong Province Natural Science Foundation under Grant ZR2025QA30 and ZR2022YQ62, the Beijing Natural Science Foundation under Grant L242011, and the Beijing Nova Program 20220484025. The work of Guanghui Wen was supported by the National Natural Science Foundation of China through Grant Nos. 62325304, U22B2046. {\it{(Corresponding author: Zhen Gao; Ziwei Wan.)}}}
        \thanks{Yi Tao and Zhuoran Li are with the School of Information and Electronics and the Advanced Research Institute of Multidisciplinary Sciences, Beijing Institute of Technology, Beijing 100081, China (e-mail: \{yitao0511, lizhuoran23\}@bit.edu.cn).

        Zhen Gao is with the State Key Laboratory of CNS/ATM and the MIIT Key Laboratory of Complex-Field Intelligent Sensing, Beijing 100081, China, also with Beijing Institute of Technology (BIT), BIT, Zhuhai 519088, China, also with the Advanced Technology Research Institute, BIT (Jinan), Jinan 250307, China, and also with the Yangtze Delta Region Academy, BIT (Jiaxing), Jiaxing 314019, China (e-mail: gaozhen16@bit.edu.cn).
        
        Ziwei Wan is with the Yangtze Delta Region Academy, Beijing Institute of Technology (Jiaxing), Jiaxing 314019, China (e-mail: ziweiwan@bit.edu.cn).
        
         Chunli Zhu is with the School of Mechatronics Engineering, Beijing Institute of Technology, Beijing 100081, China (e-mail: chunlizhu@bit.edu.cn).
 
	   Guanghui Wen is with the School of Automation, Southeast University, Nanjing 210096, China (e-mail: ghwen@seu.edu.cn).
	
        Tuan Li and Dezhi Zheng are with the Advanced Research Institute of Multidisciplinary Sciences, Beijing Institute of Technology, Beijing 100081, China (e-mail: \{6120210230, zhengdezhi\}@bit.edu.cn).

        Lei Chen is with the School of AI, Beijing Institute of Technology, Beijing 100081, China (e-mail: leichen@bit.edu.cn).
        
        Dusit Niyato is with the College of Computing and Data Science, Nanyang Technological University, Singapore 639798, (e-mail: dniyato@ntu.edu.sg).
		}
	}
	\maketitle
	
	\begin{abstract}
		The integrated sensing and communications (ISAC) can achieve the sharing of hardware and spectrum resources, enabling efficient data transmission and environmental sensing. This fusion is particularly important for unmanned aerial vehicle (UAV) swarms, as it enhances the overall performance, flexibility, and efficiency of such systems. 
		To facilitate the collaborative operations among UAVs, this paper proposes an ISAC solution based on the pseudo-random time-division multiplexing (TDM)-multiple input multiple output (MIMO) millimeter-wave (mmWave) frequency modulated continuous wave (FMCW).
		Specifically, a novel ISAC chirp waveform is proposed to modulate data in both the delay domain and complex amplitude, while also possessing high-precision sensing capabilities.
		To address challenges in the TDM-MIMO, we utilize the pseudo-random antenna selection and compressed sensing algorithms, ensuring that the maximum unambiguous velocity is not compromised. 
		Moreover, by employing a chirp-division multiple access scheme, we propose an interference-free multiple antenna transmission scheme to achieve dynamic allocation of time-frequency resources and multi-user transmission.
		Finally, we propose a communication and sensing fusion-based dynamic iterative computation scheme, simultaneously achieving data demodulation and sensing parameter estimation.
		Simulation results show that the proposed scheme can achieve ISAC under the dynamic flight scenarios of UAVs. Meanwhile, the scheme outperforms the mmWave-LoRadar in communication and sensing performance, yet its sensing performance is slightly lower than that of the traditional FMCW. Under the urban clutter modeling, the scheme still maintains favorable robustness despite a certain degree of performance degradation.
		
	\end{abstract}
	\vspace*{-1mm}
	
	\IEEEpeerreviewmaketitle

    \vspace{-3mm}
	\section{Introduction}
	Radar sensing and wireless communication, traditionally two distinct domains rooted in the utilization of electromagnetic waves, have evolved independently throughout history \cite{lhs, zhengao1,lxh, shicongliu}. 
	However, they share numerous similarities in several critical areas \cite{fanliu}.
	In light of the recent millimeter-wave (mmWave) technology advancements, these similarities present a promising opportunity for integrating these two technologies. Such integration could potentially yield synergistic gains, paving the way for a breakthrough in the next-generation network \cite{synergistic,zyf,xingyuzhou,wangkuiyu,wangyang}.

    \subsection{Related Works}
	Currently, the design of integrated sensing and communications (ISAC) systems can be categorized into three distinct paradigms: joint design, communication-centric design, and sensing-centric design. Each paradigm features unique approaches to balancing communication and sensing performance.
	For instance, the joint design facilitates a flexible trade-off between communication and sensing, unconstrained by existing standards \cite{jrc}. While this design epitomizes the optimal objective of ISAC, its realization is fraught with complexities.
	Consequently, existing communication or sensing technologies are usually given precedence. 
	For example, numerous communication-centric designs prominently feature orthogonal frequency division multiplexing (OFDM) to achieve high-speed data transmission and realize specific sensing functionalities \cite{zhengao,ranbo, preamble,IET}.
	For these designs, the data transmission rate emerges as the key performance metric, while radar sensing is treated as an additional feature. 
	
	On the other hand, the sensing-centric design can achieve optimal sensing performance, rendering it appropriate for applications that necessitate high-precision sensing \cite{FRaC, MFSK-LFMCW, LoRadar}. 
	Among sensing-centric designs, the most commonly used is the linear frequency modulation waveform \cite{chirp}, with a typical example being the mmWave frequency modulated continuous wave (FMCW) radar \cite{radar}.
	The chirp waveform is particularly well-suited for sensing-centric designs due to its superior tolerance to Doppler effects, large time-bandwidth product, and constant modulus characteristics \cite{chirp}. 
	By embedding data within the chirp waveform, it is possible to simultaneously accomplish data transmission and radar sensing. 
	For instance, in \cite{FRaC}, data modulation is accomplished through carrier selection, antenna selection, waveform permutation, and complex amplitude modulation.
	Nevertheless, the maximum likelihood scheme employed in the decoding algorithm entails high computational complexity. 
	 A LoRadar system \cite{LoRadar} based on the combination of FMCW and LoRa technologies is proposed, which generates two FMCWs with time offsets, producing harmonic signals that contain the frequency difference between the two FMCWs.
	The harmonic signals can serve as the carrier frequency of the Long Range (LoRa) signal \cite{LoRa} for data modulation.
	However, harmonic signals are weaker than the transmit signals, which reduces the communication distance. Worse still, these two functions operate in different frequency bands, failing to achieve spectrum sharing.
	
	Nowadays, unmanned aerial vehicle (UAV) swarms are emerging as a key enabling technique with their deployment flexibility and extensive coverage, showcasing immense potential in a wide range of applications \cite{uav}. 
	Unlike terrestrial communication systems, UAV swarms possess exceptional maneuverability and more convenient deployment options, facilitating rapid communication coverage in emergencies \cite{uav com}. 
	To support UAV swarm collaborative networks, each UAV must be equipped with the ability to sense the environment and communicate with the other.
	On the other hand, given the constraints of payload capacity, the installation of separate communication and sensing devices would result in increased power usage. 
	Luckily, due to the trend of hardware and waveform integration between sensing and communication, ISAC for the UAV swarm becomes possible \cite{holistic}.
	
    
    \subsection{Motivations}
	Despite the aforementioned advancements, the ISAC for the UAV swarm still faces various challenges. Considering the dynamic nature of the UAV swarm, ISAC poses higher demands on the waveform and hardware design and also requires more complex signal processing algorithms. Furthermore, the UAV swarm necessitates the efficient utilization of limited spectral resources while mitigating the mutual interference between sensing and communication \cite{joint}.
	Therefore, we propose an ISAC system that is developed from the mmWave FMCW MIMO radar, which can offer precise radar sensing  \cite{accurate} vital for the UAV navigation and obstacle avoidance.
	The large time bandwidth product of FMCW enhances anti-interference capabilities and ensures the stability and reliability of signal transmission in complex electromagnetic environments.
	Furthermore, the mmWave FMCW MIMO radar can achieve miniaturization without compromising its high performance. Particularly, by employing virtual array technology \cite{vr_m}, it can decrease the number of required antennas to reduce the hardware complexity and payload. The miniaturized radar can be easily mounted on UAVs with minimal payload and can also be seamlessly integrated with other devices.
		\vspace{-2mm}
		\subsection{Contributions}
		In this paper, we propose a mmWave ISAC system based on pseudo-random time division multiplexing (TDM)-multiple input multiple output (MIMO) FMCW. The ISAC system aims to realize the simultaneous processing of radar sensing and communication within the UAV swarm. Based on the existing FMCW architecture, minimal modifications to hardware equipment are required. Meanwhile, this system features excellent compatibility and expandability, enabling adaptation to other devices and providing potential possibilities for integration with other modulation schemes.
		
		To distinguish devices with distinct roles, we name the device in the data modulation role as the {\it{active terminal (AT)}}, while the device in the data demodulation role as the {\it{passive terminal (PT)}}. 
		It should be noted that AT and PT share identical hardware designs, therefore, AT and PT roles of UAVs can change with the subsequent tasks, demonstrating the hardware sharing in the proposed system.
		Specifically, our main contributions can be summarized as follows.
		
		\begin{itemize}
		\item{
		We propose an ISAC solution based on the mmWave pseudo-random TDM-MIMO FMCW. The proposed ISAC chirp waveform has good radar sensing capability, and is capable of transmitting information by modulating data in both the delay domain\footnote{In our proposed waveform, the distance estimated at the receiver essentially includes two parts: the true distance parameter and the communication data modulated in the delay domain. Therefore, we also refer to the data in the delay domain as the data in the {\it equivalent distance domain}.} and complex amplitude.
		}
			
		\item{
		We utilize the chirp division multiple access (Chirp-DMA) to achieve an orthogonal allocation of time-frequency (TF) resources, ensuring that ISAC between different AT-PT pairs in different resource blocks (RBs) does not interfere with each other. By combining Chirp-DMA with TDM-MIMO, we develop an interference-free multiple antenna transmission scheme, allowing AT to simultaneously utilize multiple transmit antennas for ISAC, with each antenna corresponding to a distinct RB, which enables the multi-stream ISAC with multiple PTs.}
			
		\item{In consideration of the detrimental effects of velocity ambiguity in the TDM-MIMO \cite{ambiguity}, we utilize a pseudo-random TDM-MIMO scheme coupled with compressed sensing (CS) algorithms, which can considerably enhance the maximum unambiguous velocity for radar sensing.
		Moreover, we introduce an ISAC frame structure to fully harness ISAC capabilities for the UAV swarm.	
			} 
			
		\item{We propose a data compensation-aided sensing parameter estimation (DCA-SPE) scheme for AT and a joint sensing parameter estimation and data demodulation (J-SPEDD) scheme for PT to achieve ISAC simultaneously.
		AT directly performs data removal and parameter estimation, and utilizes the extended Kalman filter (EKF) algorithm for tracking. 
		PT performs distance parameter estimation and uses EKF to predict the current distance parameters, and extracts the data modulated in the delay domain by the difference in the delay domain. 
		Then, data in the delay domain is removed, and compensation is performed based on the EKF-predicted velocity parameters. Therefore, data modulated in the complex amplitude can be obtained. Subsequently, the PT removes data in the complex amplitude and realizes the true parameter estimation, which is subsequently utilized for tracking.}
			
		\end{itemize}
		
		\emph{Notations:} Column vectors and matrices are denoted by boldface lower and upper-case symbols, respectively. $j=\sqrt{-1}$ is the imaginary unit.
		$\mathbb{C}$ is the set of complex numbers. $\Pi(\cdot)$ is the unit rectangular window function. $\left\lfloor\cdot\right\rfloor$ is the flooring function. $\odot$ is the Hadamard product. 
		$c$ is the light velocity.
		${\bf{A}}\left[\mathcal{I},:\right]$ and ${\bf{A}}\left[:,\mathcal{I}\right]$ are the sub-matrix consisting of rows and columns in matrix ${\bf{A}}$ indexed by the set $\mathcal{I}$, respectively. ${\bf{a}}\left[\mathcal{I}\right]$ is the sub-vector consisting of the elements in vector ${\bf{a}}$ indexed by $\mathcal{I}$.
		${\bf{I}}_N$ is the $N$-order identity matrix.
		$|\cdot|$ denotes the absolute value.
		${\rm{A}} \left(\cdot\right)$ is the averaging process.
		$\left[\cdot\right]^{\rm{T}}$ and $\left[\cdot\right]^{\rm *}$ are the transpose and conjugate operations, respectively. 
		\vspace{-2mm}
		\section{System Model}
		Based on the environmental characteristics of UAV swarms, this paper focuses on the collaborative mechanism within UAV swarms, namely, that the ISAC is mainly implemented among UAVs. Furthermore, given the long distances between UAVs, we adopt a far-field point target approximation to simplify the UAV modeling.
	AT performs sensing by transmitting and receiving signals independently, thus featuring a round-trip path.
        Its path loss model can be reasonably simplified to consider only the line-of-sight (LOS) path. 
        This simplification is based on the the Friis free space formula and the additional performance loss caused by each reflection.
        Consequently, the received signal power of non-line-of-sight (NLOS) paths is significantly weaker than that of the LOS path.
		In contrast, for PT, the ISAC must account for NLOS paths because NLOS paths are inherently present in real-world wireless environments. Additionally, rather than merely being interference, these NLOS paths can be beneficially utilized. For instance, through parameter estimation, PT can achieve the passive sensing of multiple targets by leveraging the multipath.
		
		\vspace{-3mm}
		\subsection{Proposed ISAC Chirp Waveform}
		We propose a novel ISAC chirp waveform derived from FMCW, where data is modulated in both the delay domain and complex amplitude. 
		The {fast-time} and {{slow-time}} correspond to the number of sampling points within a single chirp and the number of chirps within a frame, respectively. 
		$P$ chirps are transmitted in one frame, and $p \in \{0,1,\ldots,P-1\}$ denotes the slow-time index. 
		The basic ISAC chirp signal $x_p(t)$ for the $p$th chirp can be expressed as		
		\begin{align}
			\label{chirp_origin}
			\begin{aligned}		x_p(t)=\beta_{\mathrm{tx}}\left[\beta_{p}^{(\mathrm{D})}\right]^{*}\Pi\left(\frac{t-p\bar{T}}{\tilde{T}}\right)
				e^{js \left(t-p\bar{T}-\tau^{(\mathrm{D})}_{p}\right)},
			\end{aligned}
		\end{align}
		where $\beta_{\mathrm{tx}}$ is the amplitude of the transmit signal, ${\tilde{T}}$ is the time duration of a single chirp, $T_{\rm{GI}}$ is the time duration of the guard interval, and $\bar{T}=\tilde{T}+T_{\rm{GI}}$ denotes the duration of a time slot, $\beta_{p}^{(\mathrm{D})}$ and $\tau^{(\mathrm{D})}_{p}$ are the data modulated in the complex amplitude and delay domain for the $p$th chirp, respectively, and $s(t)$ is represented as
		\begin{align}
			s(t)=2\pi \left(f_c t+\alpha t^2/2 \right),
		\end{align}
		where $f_{c}$ is the carrier frequency, $\alpha=B/\tilde{T}$ is the chirp rate, and $B$ is the bandwidth. 
		\begin{figure}[t] 
			\captionsetup{font={footnotesize}, name = {Fig.}, labelsep = period}
			\centering
			\begin{subfigure}
				\centering
				\includegraphics[width=0.86\linewidth]{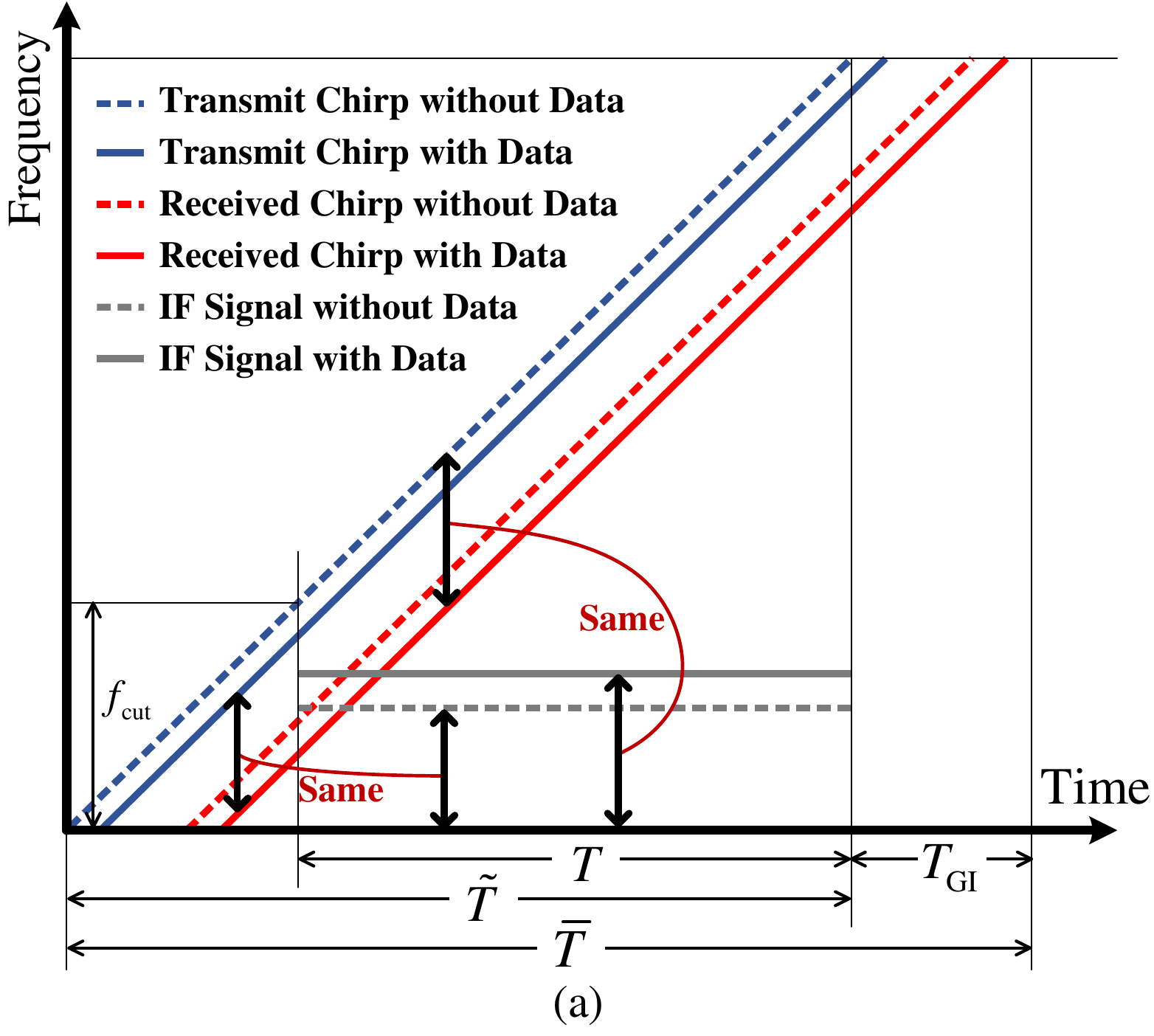} %
			\end{subfigure}			
			\begin{subfigure}
				\centering
				\includegraphics[width=0.86\linewidth]{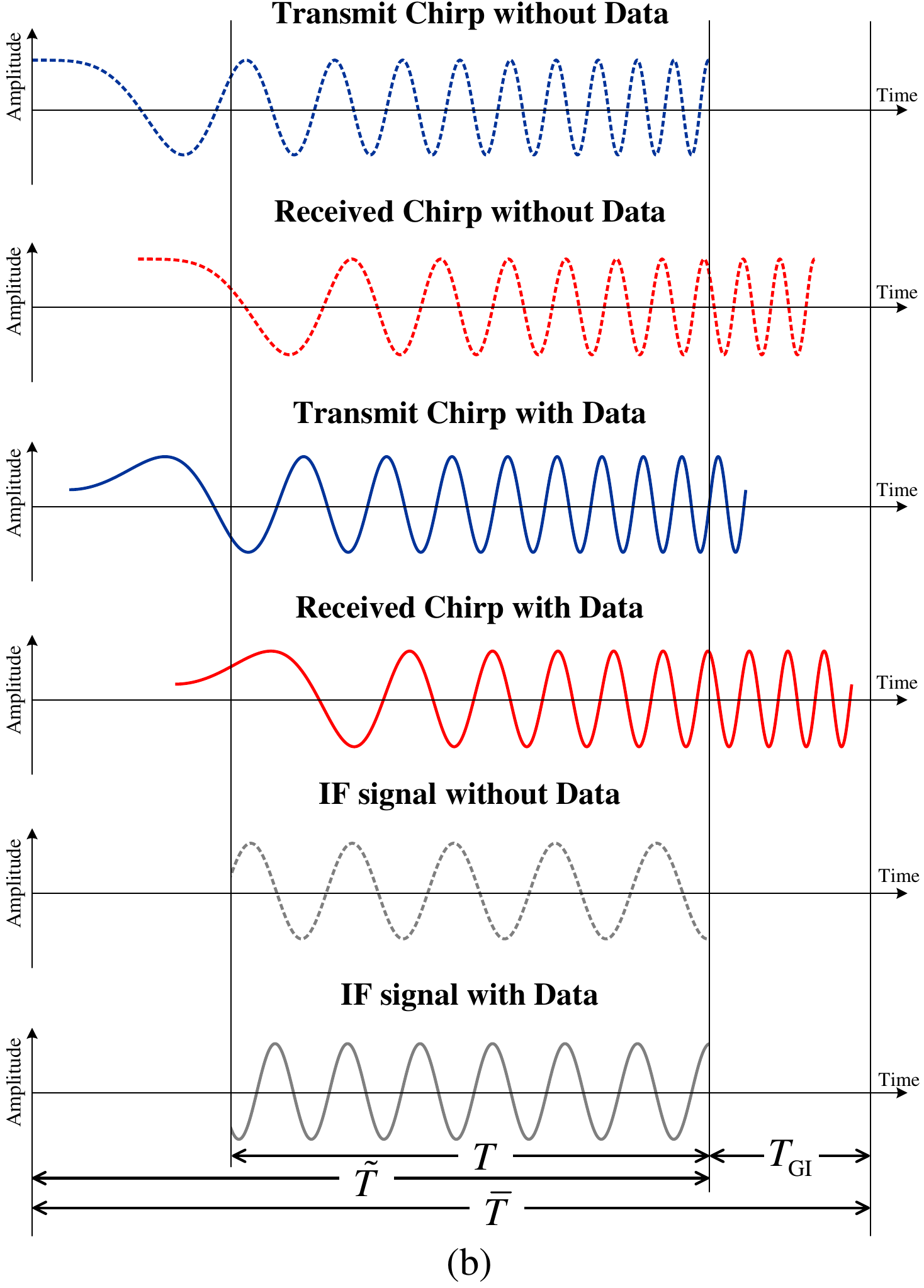} 
			\end{subfigure}
			\vspace{-2mm}
			\caption{Illustrations of the proposed ISAC chirp waveform: (a) time-frequency domain diagrams; (b) time domain diagrams of the real parts.}
			\vspace{-5mm}
		\end{figure}
		Fig. 1 illustrates the design of the ISAC chirp waveform in both the time-frequency domain and time domain, respectively, presenting the waveform with modulated data and the unmodulated waveform (i.e., the conventional chirp with $\beta_{p}^{(\mathrm{D})}=1$ and $\tau^{(\mathrm{D})}_{p}=0$), along with their respective received signals and intermediate frequency (IF) signals.
		It is noteworthy that the ISAC signal after data modulation retains the intrinsic linear frequency modulation characteristics, which ensures its compatibility with most existing parameter estimation algorithms designed for FMCW.
		However, given the superposition of complex amplitude variations and delays, customized ISAC solutions are required to enable demodulation and parameter estimation.
		
		\vspace{-3mm}
		\subsection{Proposed ISAC Antenna Architecture}
		In UAV applications, angles exhibit two-dimensional characteristics. 
		For the angle estimation, we employ the L-shaped uniform linear arrays (ULAs)\footnote{For the vast majority of FMCW applications, mutual coupling between densely packed antennas and antenna calibration requirements for coherent processing are generally not regarded as a core limiting factor.} \cite{L-SHAPE}. 
		Compared to the uniform planar array (UPA), L-shaped ULAs present a more streamlined configuration, effectively decoupling the two-dimensional problem into two independent one-dimensional problems, significantly reducing the complexity of signal processing.
		Additionally, by utilizing virtual array technology \cite{vr_m}, L-shaped ULAs can achieve an array aperture equivalent to that of UPAs with fewer antennas, which aligns with the requirement of lightweight hardware and high-precision sensing for UAV swarm collaboration.
		
		\begin{figure}
			\captionsetup{font={footnotesize}, name = {Fig.}, labelsep = period}
			\centering
			\includegraphics[width=8.6cm, keepaspectratio]
			{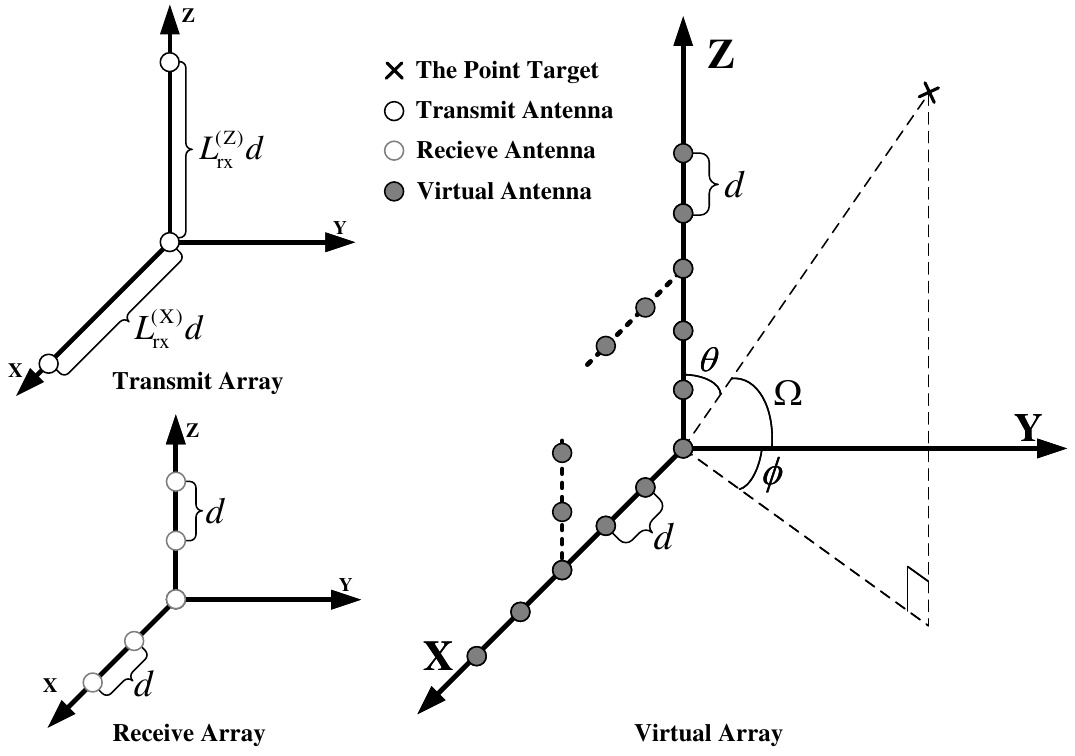}
			\caption{The schematic diagram of the L-shaped ULAs architecture, where $L_{\rm{tx}}=3$, $L_{\rm{rx}}=5$, $L_{\rm{tx}}^{(\rm{X})}=L_{\rm{tx}}^{(\rm{Z})}=2$, and $L_{\rm{rx}}^{(\rm{X})}=L_{\rm{rx}}^{(\rm{Z})}=3$.}
			\vspace{-6mm} 
		\end{figure}
		The schematic diagram of the L-shaped ULAs architecture deployed on each UAV is depicted in Fig. 2. 
		Each UAV is equipped with a total of $L_{\rm{tx}}$ transmit antennas and $L_{\rm{rx}}$ receive antennas. 
		On one of the axis, there are $L_{\rm{tx}}^{(\triangle)}$ transmit antennas and $L_{\rm{rx}}^{(\triangle)}$ receive antennas, where $(\triangle)$ can be replaced by $\rm{X}$ or $\rm{Z}$ to denote the horizontal or vertical ULA, respectively.
		The index of transmit and receive antennas in one axis are respectively denoted as $l^{(\triangle)}_{\rm{tx}} \in \{0,1,\ldots, L^{(\triangle)}_{\rm{tx}}-1\}$ and $l^{(\triangle)}_{\rm{rx}} \in \{0,1,\ldots, L^{(\triangle)}_{\rm{rx}}-1\}$.
		The index of all transmit and receive antennas are denoted as $l_{\rm{tx}} \in \{0,1,\ldots, L_{\rm{tx}}-1\}$ and $l_{\rm{rx}} \in \{0,1,\ldots, L_{\rm{rx}}-1\}$, respectively. 
		The antenna farthest from the origin in the $x$-axis is defined as $l_{\rm{tx}}=0$ or $l_{\rm{rx}}=0$ for convenience. However, when considering the individual axis, the antenna located at the origin is regarded as $l^{(\triangle)}_{\rm{tx}}=0$ or $l^{(\triangle)}_{\rm{rx}}=0$.
		Consequently, there is a correspondence between $l_{\rm{tx}}^{(\triangle)}$ and $l_{\rm{tx}}$, as well as between $l_{\rm{rx}}^{(\triangle)}$ and $l_{\rm{rx}}$. 
		The spacing between adjacent receive and transmit antennas is $d$ and $L_{\rm{rx}}^{(\triangle)} d$, respectively, therefore, the spacing between adjacent virtual antennas is $d$. 
		The total number of the virtual antennas is $L=L_{\rm{tx}}L_{\rm{rx}}$.
		As depicted in Fig. 2, the projected azimuth angle is defined as $\phi $, the azimuth angle as $\Omega$, and the elevation angle as $\theta$. These angles are related by $\cos \Omega = \sin \theta \cos \phi $. 
		In this paper, we focus on obtaining $\theta$ and $ \phi $.
		
		\vspace{-3mm}
		\subsection{Interference-Free Multiple Antenna Transmission with Pseudo-Random TDM-MIMO} 
		To achieve interference-free transmission and improve the transmission rate, we introduce the Chirp-DMA \cite{fanliu} to optimize the time-frequency (TF) resource allocation.
		For FMCW, the maximum unambiguous delay is given by $T_d = f_s T/B$, where $f_s$ denotes the sampling frequency and $T$ represents the effective sampling time interval, as illustrated in Figs. 1 and 3. Based on the principle of time-staggered FMCW \cite{time-staggered}, if the time intervals between the start times of different transmit signals all exceed $2T_d$, and target distances are strictly confined within the maximum unambiguous distance, the IF signals corresponding to each transmit signal can be effectively separated using respective low-pass filters (LPFs) with a cutoff frequency set to $f_{\rm{cut}} = f_s$.
		Theoretically, TF resources can be subdivided into $\lfloor \bar{T}/(2 T_d) \rfloor$ resource blocks (RBs), which can be allocated to achieve orthogonal transmission.
		As illustrated in Fig. 3, each RB has a base length of $2 T_d$, a height of $B$, and a slope of the diagonal $\alpha$. $P$ chirps within one frame in \eqref{chirp_origin} occupy the RBs having the same color. Since the delay of received signals is confined within $T_d$, there is no mutual interference between different RBs.
        A typical mixing case is shown in Fig. 4. Taking the transmit signal of RB $1$ as an example, although it undergoes frequency mixing with the received signals of two RBs, as long as a spacing between them is maintained and an appropriate LPF is selected, we can obtain only the result of frequency mixing with the received signal in RB $1$. This ultimately avoids interference between signals.
        In addition, the interval setting of $2T_d$ can further tolerate a certain degree of synchronization errors, mitigate the risk of non-orthogonality among IF signals under non-ideal synchronization, and thereby enhance the system robustness.
		
		\begin{figure}
			\captionsetup{font={footnotesize}, name = {Fig.}, labelsep = period}
			\centering
			\includegraphics[width=8.6cm, keepaspectratio]
			{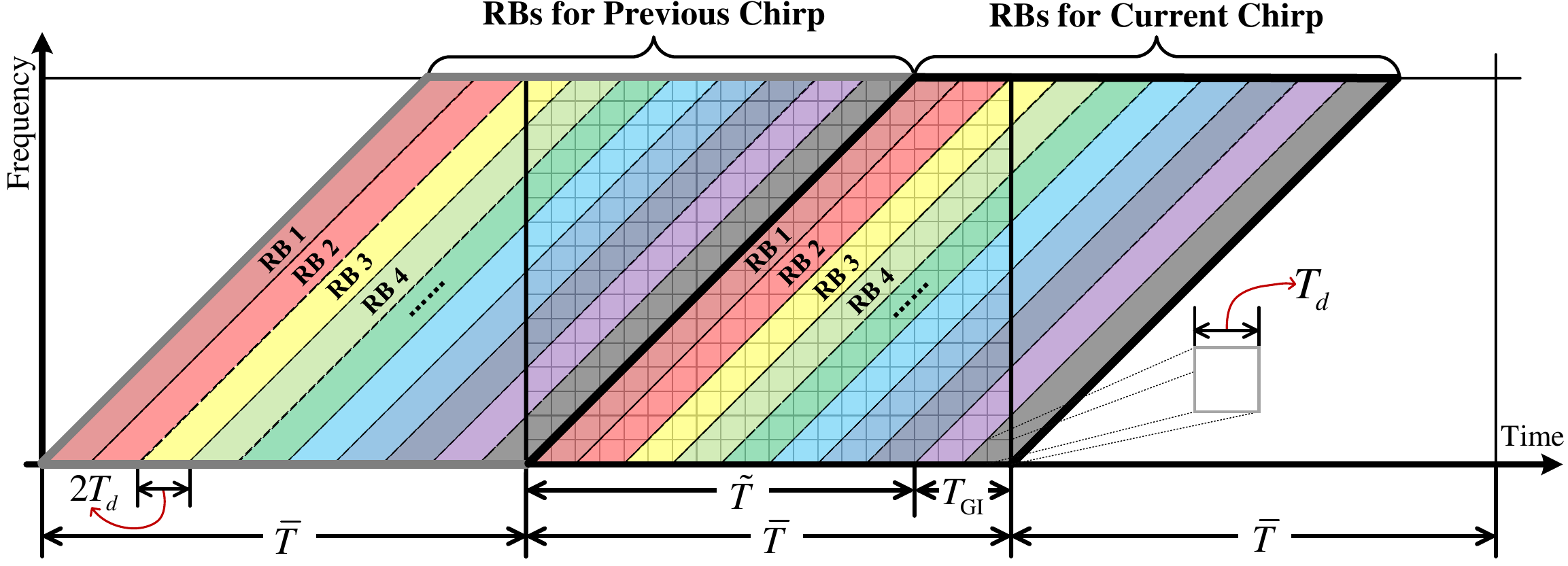}
			\caption{The schematic diagram of the Chirp-DMA in the time-frequency domain.} 
			\vspace{-7mm}
		\end{figure}

        	\begin{figure}
			\captionsetup{font={footnotesize}, name = {Fig.}, labelsep = period}
			\centering
			\includegraphics[width=8.6cm, keepaspectratio]
			{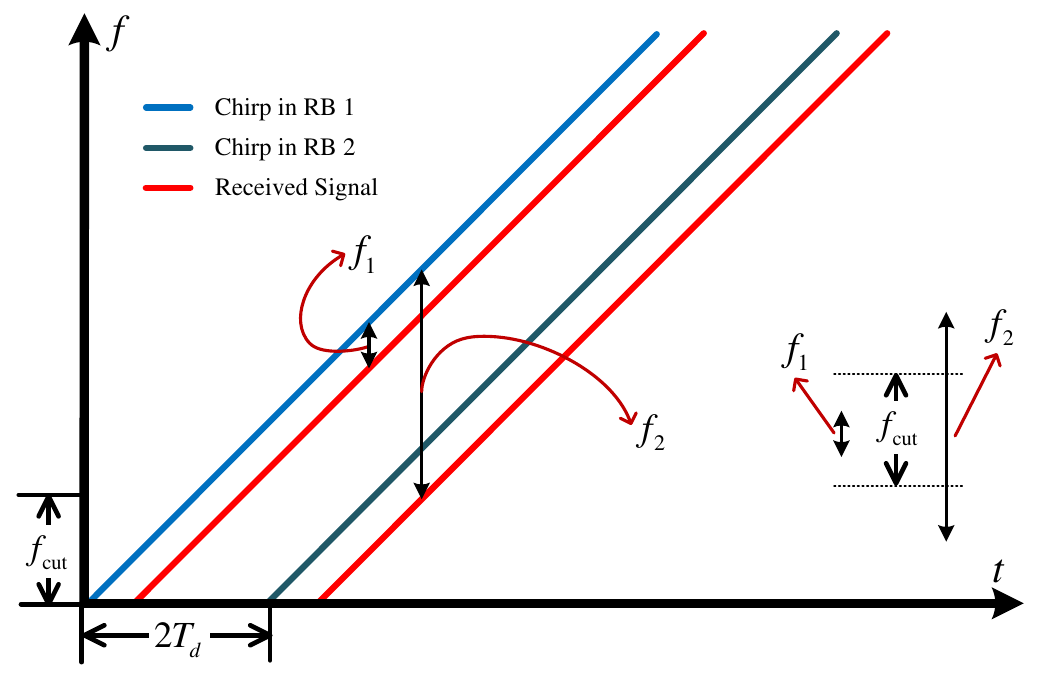}
			\caption{Schematic diagram of the principle for Chirp-DMA to achieve orthogonality using LPF.} 
			\vspace{-7mm}
		\end{figure}
		Leveraging the Chirp-DMA and TDM-MIMO, we propose an interference-free multiple antenna transmission scheme. 
		For parameter estimation, the receive antennas must distinguish signals coming from different transmit antennas. The conventional TDM-MIMO scheme achieves orthogonality by allowing different antennas to transmit during non-overlapping time intervals, as illustrated in Fig. 5(a).
		Specifically, each frame consists of multiple cycles, with each cycle containing $L_{\rm{tx}}$ time slots. Within a single cycle, each time slot is occupied by one of the $L_{\rm{tx}}$ transmit antennas.
		\begin{figure*}
			\captionsetup{font={footnotesize}, name = {Fig.}, labelsep = period}
			\centering
			\includegraphics[width=17.5cm, keepaspectratio]
			{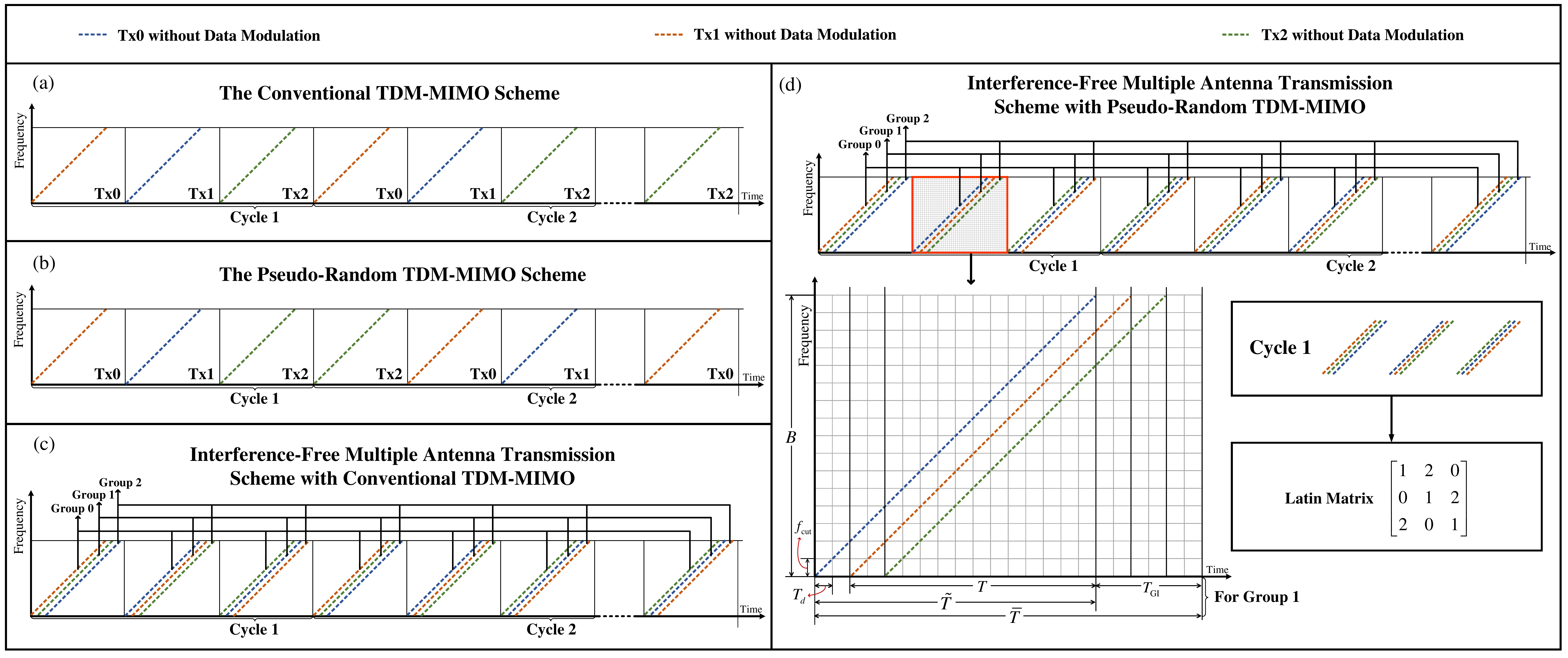}
			\caption{An illustration of the design evolution of  the interference-free multiple antenna transmission scheme with pseudo-random TDM-MIMO when $ L^{(\rm{X})}_{\mathrm{tx}}  = 2 $ and $ L^{(\rm{Z})}_{\mathrm{tx}}  = 2 $: (a) the conventional TDM-MIMO scheme; (b) the pseudo-random TDM-MIMO scheme; (c) the interference-free multiple antenna transmission scheme with conventional TDM-MIMO; (d) the interference-free multiple antenna transmission scheme with pseudo-random TDM-MIMO.} 
			\vspace{-4mm}
		\end{figure*}
		We have made targeted improvements to the conventional TDM-MIMO scheme. Specifically, based on the Chirp-DMA, AT can select multiple RBs within each time slot.
		In Fig. 5(c), from the perspective of distinct RBs, the transmission exhibits characteristics of the TDM-MIMO. 
		Specifically, signals can be divided into $M$ groups, with each group corresponding to different RBs, and $M \leq L_{\mathrm{tx}}$. $m \in \{0,1,\ldots,M-1\}$ is the index for the chirp group. For simplicity, we define $ M = L_{\mathrm{tx}} $, implying that all transmit antennas are utilized for ISAC.
		From the perspective of a single time slot, all transmit antennas are used nearly simultaneously. 
		Therefore, $ L_{\mathrm{tx}} $ transmit antennas can allocate up to $ L_{\mathrm{tx}} $ RBs to achieve ISAC. 
		
		However, the mechanism of TDM-MIMO results in a reduction in the unambiguous velocity \cite{chirp}.
		Therefore, inspired by \cite{Random}, we propose a pseudo-random TDM-MIMO scheme\footnote{The analysis of pseudo-random TDM-MIMO can be decomposed into the bounds of parameter estimation and the analysis of CS algorithms, which will be elaborated in detail in subsequent sections.} for ambiguity-free velocity estimation.
		The pseudo-random TDM-MIMO scheme is also shown in Fig. 5(b). 
		Within one cycle, each group comprises $L_{\rm{tx}}$ transmit antennas that are activated in a pseudo-random order. 
		Furthermore, by combining the interference-free multi-antenna transmission scheme, we derive a transmission scheme suitable for the ISAC waveform. A typical example is illustrated in Fig. 5(d), where $ L_{\mathrm{tx}}  = 3$, $ {L}^{(\rm{X})}_{\mathrm{tx}}  = 2 $, $ L^{(\rm{Z})}_{\mathrm{tx}}  = 2 $, and $ T_{\rm{GI}} = 2 L_{\mathrm{tx}} T_d $.
		In one cycle, if different RBs are taken as rows and groups as columns, the antennas used within a cycle exhibit the form of a Latin matrix, which is a stricter constraint for antenna selection.
		However, the pseudo-random TDM-MIMO renders the discrete Fourier transform (DFT) unusable. To address this issue, we introduce CS algorithms. By utilizing CS, it is able to achieve ambiguity-free velocity estimation and eliminate phase errors induced by target motion when estimating angles \cite{ambiguity} in the TDM-MIMO. 
		\begin{figure*}
			\centering
			\captionsetup{font={footnotesize}, name = {Fig.}, labelsep = period}
			\includegraphics[width=17cm, keepaspectratio]
			{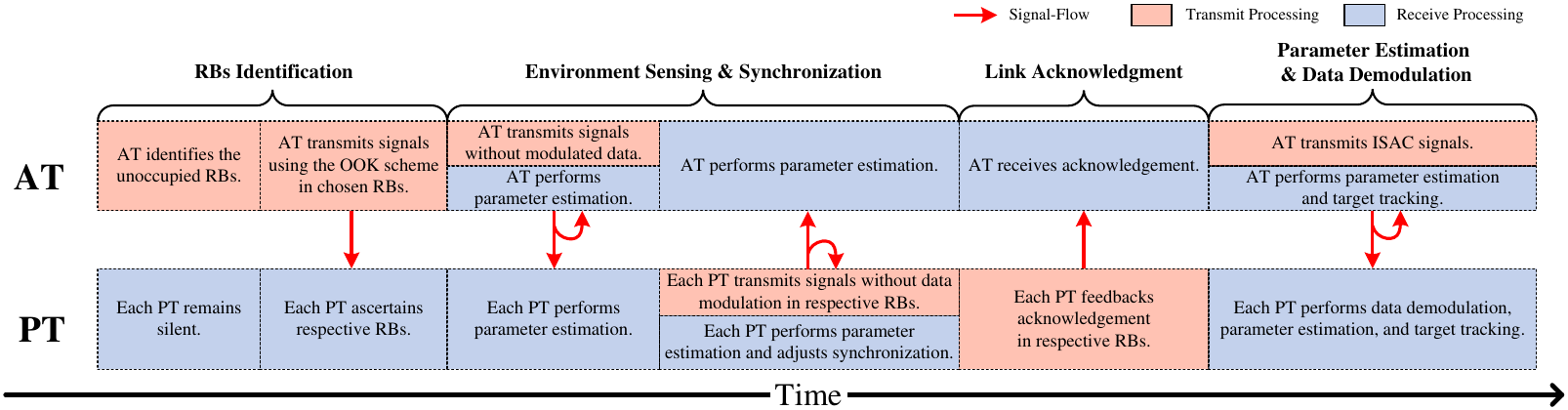}
			\caption{The frame structure of the proposed ISAC system for the UAV swarm.} 
			\vspace{-7mm}
		\end{figure*}
		\vspace{-4mm}
		\section{Proposed ISAC Frame Structure}
		Building on the above ISAC design, we introduce a frame structure, which can be divided into four stages. 
		For illustration, we take a single AT as an example. In practical applications, relying on the number of RBs, the system can support multiple ATs and multiple PTs to carry out ISAC operations simultaneously.
		A simplified flowchart of the frame structure is depicted in Fig. 7.
		\begin{itemize}
			\item[1)]{{\textit{RBs Identification:}}  
				AT and PTs employ their respective procedures to determine the available RBs. Initially, utilizing the Chirp-DMA, AT identifies the currently unoccupied RBs by mixing with the environmental received signals. Following this, AT can select $M$ consecutive unoccupied RBs and transmit ISAC signals without data modulation using an on-off keying (OOK) scheme\footnote{In the OOK scheme, the transmission of a chirp is defined as a ``$1$'', while the absence of transmission is defined as a ``$0$''. } within these chosen RBs. 
				Then, by transmitting predetermined synchronization information using the OOK scheme, each PT mixes signals to ascertain the RBs that AT intends to use.}
			
			\item[2)]{{\textit{Environment Sensing \& Synchronization:}} 
				In the first round of parameter estimation, AT transmits unmodulated signals across $M$ RBs, and signals are received and processed by AT itself and each PT. After receiving signals, all terminals accomplish the first round of parameter estimation.
				Subsequently, AT enters a silent state, while PTs transmit unmodulated signals within their respective RBs. These signals are also received and processed by AT and corresponding PTs, facilitating the second round of parameter estimation. 
				By comparing the parameter estimates from two rounds, AT and PTs can obtain the synchronization delay\footnote{In this paper, we assume that perfect synchronization can be achieved.
                If synchronization errors still exist, they will introduce a fixed synchronization delay to PT, leading to a fixed deviation in sensing. Yet because the synchronization delay is constant, the variation of distance delay within a single frame is minimal, and the distance modulated data is much larger than this variation, data demodulation remains largely unaffected. However, the sum of the actual distance delay and synchronization delay is treated as the distance delay. Through a closed-loop compensation between AT and PTs (e.g., by transmitting timing correction parameters), perfect synchronization can still be achieved ultimately.}. Furthermore, if synchronization between AT and PTs is achieved through an external clock source, this stage can be omitted.
			}
			
			\item[3)]{{\textit{Link Acknowledgment:}} 
				After synchronization, each PT utilizes the OOK scheme to transmit signals in their respective RBs for acknowledgment, informing AT of the ISAC link establishment.}
			
			\item[4)]{{\textit{Parameter Estimation \& Data Demodulation:}} 
				AT transmits $M$ groups of ISAC signals carrying communication data and subsequently receives the echo signals for parameter estimation and target tracking. PTs receive the ISAC signals transmitted by AT and utilize information from previous frames to perform EKF-based parameter prediction, which enables the demodulation of data in the delay domain and complex amplitude. Then, parameter estimation and target tracking can be performed after compensating for the modulated data. Thereafter, AT and PTs repeat this process, and AT periodically receives feedback from PTs for error updates.}
		\end{itemize}

		\vspace{-2mm}
		\section{Multiple Antenna ISAC Signal Transmission Model with Pseudo-Random TDM-MIMO}
		According to \eqref{chirp_origin}, the transmit ISAC chirp signal of the $m$th group, denoted by $x_{m,p}(t)$, can be rewritten as
		\begin{align}
			\begin{aligned}
				x_{m,p}(t)=\beta_{\mathrm{tx}}\left[\beta_{m,p}^{(\mathrm{D})}\right]^{*}\Pi\left(\frac{t_{m,p}}{\tilde{T}}\right)e^{js \left(t_{m,p}-\tau^{(\mathrm{D})}_{m,p} \right)},
			\end{aligned}
		\end{align}
		where $t_{m,p} = t-p\bar{T}-2 m T_d$, $\beta_{m,p}^{(\mathrm{D})}$ and $\tau^{(\mathrm{D})}_{m,p}$ are the data modulated in the complex amplitude and delay domain for the $m$th group and the $p$th chirp, respectively.
		The number of targets is defined as $K^{(\mathrm{\cdot})}$, where $(\cdot)$ can be replaced by $(\rm{A})$ or $(\rm{P})$ to represent the process at AT or PT, respectively.
		After AT transmits the ISAC signal, ${x}_{m,p}(t)$ can be received by AT or PT.
		The received signal for the $l_{\mathrm{rx}}$th receive antenna, termed as ${\hat{x}}^{(\cdot)}_{m,p,l_{\mathrm{rx}}}(t)$, can be expressed as
		\begin{align}
			\label{re}
			\begin{aligned}
				{\hat{x}}^{(\cdot)}_{m,p,l_{\mathrm{rx}}}(t)&=\sum_{k^{(\mathrm{\rm{\cdot}})}=1}^{K^{(\mathrm{\cdot})}}\beta^{(\mathrm{\rm{\cdot}})}_{\mathrm{rx},k} \left[\beta_{m,p}^{(\mathrm{D})}\right]^{*} \Pi\left(\frac{t_{m,p}-\tau^{(\cdot)}_k(t)}{\tilde{T}}\right)\\
				&e^{js \left( t_{m,p}-\tau^{(\mathrm{D})}_{m,p} -\tau^{(\cdot)}_k(t) \right)}\\
				&e^{-j2\pi\frac{f_c}{c} l^{(\triangle)}_{m,p} L^{(\triangle)}_{\mathrm{rx}} \sin\chi^{(\mathrm{\rm{\cdot}})}_{{\rm{tx}},k}}e^{-j2\pi\frac{f_c}{c} l^{(\triangle)}_{\mathrm{rx}}\sin\chi^{(\mathrm{\rm{\cdot}})}_{{\rm{rx}},k}},
			\end{aligned}
		\end{align}
		where $k^{(\mathrm{\rm{\cdot}})} \in \{1,2,\ldots, K^{(\mathrm{\rm{\cdot}})}\}$ is the index of targets, $\beta^{(\mathrm{\rm{\cdot}})}_{\mathrm{rx},k}$ and $\tau^{(\cdot)}_k(t)$ are the received signal's amplitude and delay of the $k^{(\mathrm{\rm{\cdot}})}$th target, respectively.
		$l^{(\triangle)}_{m,p} = \left\{ l^{(\triangle)}_{\rm{tx}} \mid {\bf{E}}_m\left[{i,l_{\rm{tx}}}\right] = p \right\} \in  \{ 0,1,\ldots,L^{(\triangle)}_{\mathrm{tx}}-1 \}$ is the index of the activated transmit antenna for the $m$th group and the $p$th chirp, where the $l_{\rm{tx}}$th column of ${\bf{E}}_m \in \mathbb{C}^{(P/L_{\rm{tx}}) \times L_{\rm{tx}}}$ represents the slow-time indices of the $l_{\rm{tx}}$th transmit antenna.
		${\chi}^{(\cdot)}_{{\rm{tx}},k}$ and ${\chi}^{(\cdot)}_{{\rm{rx}},k}$ denote the angles of departure and arrival for the $k$th target, respectively, where $\chi$ can be replaced with $\theta$ or $\phi$ based on the corresponding relationship of ${(\triangle)}$ in $l^{(\triangle)}_{m,p}$.
		
		Then, the received signal \eqref{re} is mixed with the transmit signal without modulated data, termed as ${{x}}'_{m,p}(t)$, i.e., $\beta_{m,p}^{(\mathrm{D})}=1$ and $\tau^{(\mathrm{D})}_{m,p} = 0$ for ${{x}}_{m,p}(t)$, to generate the IF signal. 
		In the $p$th time slot for the $m$th group, we only take the time interval from $\tilde{T}-T+p\bar{T}+2mT_d$ to $\tilde{T}+p\bar{T}+2mT_d$ for ISI prevention. Therefore, the IF signal can be represented as\footnote{Notably, the velocity estimation essentially involves deriving the motion state of the target based on the phase difference variation between adjacent pulses \cite{chirp}. 
        The core prerequisite for this characteristic is that the Doppler frequency shift is much smaller than the signal bandwidth. As a spread-spectrum signal, FMCW possesses a wide bandwidth characteristic \cite{radar}. Thus, the time-frequency domain stretching or compression of FMCW caused by the Doppler effect is negligible.}
		\begin{align}
			\label{equ:y_d}
			\begin{aligned}
				y_{m,p,l_{\mathrm{rx}}}^{(\cdot)}(t) &={{x}}'_{m,p}(t) \odot \left[{\hat{x}}^{ {(\cdot)}}_{m,p,l_{\mathrm{rx}}}(t)\right]^{*}\\
				&=\sum_{k^{(\mathrm{\rm{\cdot}})}=1}^{K^{(\mathrm{\cdot})}}\tilde{\beta}^{(\cdot)}_k \beta_{m,p}^{(\mathrm{D})}  \Pi\left(\frac{t_{m,p}-(\tilde{T}-T)}{T}\right)  \\
				&e^{js(t_{m,p}) - js\left( t_{m,p}-\tau^{(\cdot)}_k(t)-\tau^{(\mathrm{D})}_{m,p} \right)}\\
				&e^{j2\pi\frac{f_c}{c} l^{(\triangle)}_{m,p}L^{(\triangle)}_{\mathrm{rx}}\sin\chi^{(\mathrm{\rm{\cdot}})}_{{\rm{tx}},k}}e^{j2\pi\frac{f_c}{c} l^{(\triangle)}_{\mathrm{rx}}\sin\chi^{(\mathrm{\rm{\cdot}})}_{{\rm{rx}},k}},
			\end{aligned}
		\end{align}		 
		where $\tilde{\beta}^{(\cdot)}_k$ is the amplitude of the received signal.
		It is worth noting that \eqref{equ:y_d} is nearly identical to the parameter estimation formula for FMCW, with the key difference lying in the additional data modulation component and $m$.

		\subsection{Received IF ISAC Signal after Mixing at AT}
		For AT, $\tau^{(\rm{A})}_k (t)$ can be expressed as 
		\begin{align}
			\label{akt}
			\tau^{(\rm{A})}_k (t) = 2\left(d^{(\rm{A})}_k + v^{(\rm{A})}_kt\right)/c,
		\end{align}
		where $d^{(\rm{A})}_k$ and $v^{(\rm{A})}_k$ denote the distance and relative radial velocity between AT and the $k^{(\rm{A})}$th target, respectively. $2$ in \eqref{akt} is due to the round-trip path.
		The self-transmitting and self-receiving characteristic of AT leads to $\theta^{(\rm{A})}_{{\rm{tx}},k} = \theta^{(\rm{A})}_{{\rm{rx}},k}$ and $\phi^{(\rm{A})}_{{\rm{tx}},k} = \phi^{(\rm{A})}_{{\rm{rx}},k}$.
		Consequently, \eqref{equ:y_d} for AT is rewritten as
		\begin{align}
			\label{equ:y_a}
			\begin{aligned}
				y_{m,p,l_{\mathrm{rx}}}^{(\rm{A})}(t)&{\approx}\sum_{k^{(\rm{A})}=1}^{K^{(\rm{A})}}\tilde{\beta}_k^{(\rm{A})} \beta_{m,p}^{(\mathrm{D})} \Pi\left(\frac{t_{m,p}-(\tilde{T}-T)}{T}\right) \\
				&e^{j2\pi\left(f_{c}+\alpha t_{m,p}\right) \frac{2d^{(\rm{A})}_{k}+d_{m,p}^{\mathrm{(D)}}+2v^{(\rm{A})}_{k} t}{c}}\\
				&e^{j2\pi\frac{f_c}{c} \left(l^{(\triangle)}_{m,p}L^{(\triangle)}_{\mathrm{rx}} + l_{\mathrm{rx}}\right)\sin\chi^{(\rm{A})}_{{\rm{tx}},k}},
			\end{aligned}
		\end{align}
		where $d_{m,p}^{(\mathrm{D})} = c \tau_{m,p}^{(\mathrm{D})}$ is the equivalent distance corresponding to $\tau_{m,p}^{(\mathrm{D})}$ modulated in the delay domain for the $p$th chirp and the $m$th group.
		In \eqref{equ:y_a}, the approximation is because $\alpha \left( \tau_k^{(\rm{A})}(t)+\tau_{m,p}^{(\mathrm{D})}\right) \ll f_{c}$ and $\tau^{(\rm{A})}_k(t)+\tau_{m,p}^{(\mathrm{D})} \ll t$.
		
		Then, all IF signals are digitized using analog-to-digital converters (ADCs). Given a sampling frequency $f_s$ and $t=n {T_s}+p\bar{T}+2 m T_d+\tilde{T}-T$, where $n \in \{0,1,\ldots,f_{s}T-1\}$ is the index of sampling points for one time slot.
		${T_s}= 1/{f_s}$ is defined as the sampling interval, and $N = f_s T$ is the number of sampling points in one time slot. 
		For AT, the sampled IF signal for the $m$th group and the ${l_{\mathrm{rx}}}$th receive antenna, termed as ${\mathbf{Y}_{m,{l_{\mathrm{rx}}}}^{(\rm{A})}} \in \mathbb{C}^{N \times P}$, can be represented as
		\begin{align}
			\label{y_d_A}
			\begin{aligned}
				{\mathbf{Y}_{m,{l_{\mathrm{rx}}}}^{(\rm{A})}}[n,p]
				&\stackrel{\Delta}{=}y_{m,p,l_{\mathrm{rx}}}^{(\rm{A})}(t)|_{t=nT_{s}+p\bar{T}+2 m T_d+\tilde{T}-T}
				\\&\approx \sum_{k^{(\rm{A})}=1}^{K^{(\rm{A})}} {\beta}_k^{(\rm{A})}\beta_{m,p}^{(\mathrm{D})} e^{j2\pi \frac{n}{N} \left(f_{d,k}^{{(\rm{A})}}+f_{d,m,p}^{{(\rm{D})}}\right)}
				\\&e^{j2\pi \frac{p}{P}f_{v,k}^{{(\rm{A})}}}
				e^{j2\pi\frac{f_c}{c} \left(l^{(\triangle)}_{m,p}L^{(\triangle)}_{\mathrm{rx}} + l^{(\triangle)}_{\mathrm{rx}}\right)\sin\chi^{(\rm{A})}_{{\rm{tx}},k}},
			\end{aligned}
		\end{align}
		where $\beta^{(\rm{A})}_k$ is the amplitude of the received signal for AT after sampling.
		The approximation in \eqref{y_d_A} is because $ v_k^{(\rm{A})} 2 mT_d  \ll d_k^{(\rm{A})}$, $\alpha n T_s  \ll  f_c$, and $\alpha \left(\tilde{T}-T \right)  \ll  f_c$. $f_{d,k}^{{(\rm{A})}}$ and $f_{v,k}^{(\rm{A})}$ denote the normalized distance frequency and normalized Doppler frequency for AT, respectively represented as
		\begin{align}
			\label{fadvk}
			\begin{aligned}
				f_{d,k}^{{(\rm{A})}}&=2N T_s \alpha d_{k}^{(\rm{A})} /c \in[0,N),\\ f_{v,k}^{(\rm{A})}&=2P\bar{T}f_{c}v_{k}^{(\rm{A})}/c\in[0,P), 
			\end{aligned}
		\end{align}	
		and $f_{d,m,p}^{(\mathrm{D})}$ corresponds to the data modulated in the equivalent distance domain, represented as  
		\begin{align}
			\label{fdmp}
			f_{d,m,p}^{(\mathrm{D})}=NT_{s}\alpha d_{m,p}^{(\mathrm{D})}/c \in[0,N).
		\end{align}	
		\vspace{-6mm}
		\subsection{Received IF ISAC Signal after Mixing at PT}		
		For PT, $\tau^{(\rm{P})}_k (t)$ can be similarly expressed as 
		\begin{align}
			\tau^{(\rm{P})}_k (t)= \left(d^{(\rm{P})}_k+v^{(\rm{P})}_k t \right)/c,
		\end{align}
		where $d^{(\rm{P})}_k = d^{(\rm{P})}_{{\rm{AT}},k} + d^{(\rm{P})}_{{\rm{TP}},k}$ for the NLOS path or $d^{(\rm{P})}_k = d^{(\rm{P})}_{{\rm{AP}},k}$ for the LOS path. 
		$d^{(\rm{P})}_{{\rm{AT}},k}$, $d^{(\rm{P})}_{{\rm{TP}},k}$, and $d^{(\rm{P})}_{{\rm{AP}},k}$ are the distances from AT to the $k^{(\rm{P})}$th target, from the $k^{(\rm{P})}$th target to PT, and from AT to PT, respectively. $v^{(\rm{P})}_k = {\rm{d}}d^{(\rm{P})}_k / {\rm{d}}t$ is the velocity of the $k^{(\rm{P})}$th target. 
		Therefore, \eqref{equ:y_d} for PT is 
		\begin{align}
			\label{equ:y_p}
			\begin{aligned}
				y_{m,p,l_{\mathrm{rx}}}^{(\rm{P})}(t)&{\approx}\sum_{k^{(\rm{P})}=1}^{K^{(\rm{P})}}\tilde{\beta}_k^{(\rm{P})}\beta_{m,p}^{(\mathrm{D})}\Pi\left(\frac{t_{m,p}-(\tilde{T}-T)}{T}\right) \\
				&e^{j2\pi\left(f_{c}+\alpha t_{m,p}\right) \frac{d^{(\rm{P})}_k+d_{m,p}^{\mathrm{(D)}}+ v^{(\rm{P})}_{k}t }{c}}\\
				&e^{j2\pi\frac{f_c}{c} l^{(\triangle)}_{m,p}L^{(\triangle)}_{\mathrm{rx}}\sin\chi^{(\rm{P})}_{{\rm{tx}},k}}e^{j2\pi\frac{f_c}{c} l^{(\triangle)}_{\mathrm{rx}}\sin\chi^{(\rm{P})}_{{\rm{rx}},k}},
			\end{aligned}
		\end{align}
		where the approximation in \eqref{equ:y_p} is also similar to that in AT because $\alpha \left(\tau_k^{(\rm{P})}(t)+\tau_{m,p}^{(\mathrm{D})} \right)\ll f_{c}$ and $\tau^{(\rm{P})}_k(t)+\tau_{m,p}^{(\mathrm{D})}\ll t$.
		For PT, $\theta^{(\rm{A})}_{{\rm{tx}},k} \neq \theta^{(\rm{A})}_{{\rm{rx}},k}$ and $\phi^{(\rm{A})}_{{\rm{tx}},k} \neq \phi^{(\rm{A})}_{{\rm{rx}},k}$. Therefore, the sampled IF signal for the $m$th group and the ${l_{\mathrm{rx}}}$th receive antenna ${\mathbf{Y}_{m,l_{\mathrm{rx}}}^{(\rm{P})}} \in \mathbb{C}^{N \times P}$ is given as
		\begin{align}
			\label{y_d_P}
			\begin{aligned}
				{\mathbf{Y}_{m,l_{\mathrm{rx}}}^{(\rm{P})}}[n,p]
				\approx& \sum_{k^{(\rm{P})}=1}^{K^{(\rm{P})}} \beta^{(\rm{P})}_k \beta_{m,p}^{(\mathrm{D})} e^{j2\pi \frac{n}{N} \left(f_{d,k}^{{(\rm{P})}}+f_{d,m,p}^{{(\rm{D})}}\right)}
				\\&e^{j2\pi \frac{p}{P}f_{v,k}^{{(\rm{P})}}}
				\\&e^{j2\pi\frac{f_c}{c} l^{(\triangle)}_{m,p}L^{(\triangle)}_{\mathrm{rx}}\sin\chi^{(\rm{P})}_{{\rm{tx}},k}}e^{j2\pi\frac{f_c}{c} l^{(\triangle)}_{\mathrm{rx}}\sin\chi^{(\rm{P})}_{{\rm{rx}},k}},
			\end{aligned}
		\end{align}
		where $\beta^{(\rm{P})}_k $ represents the amplitude of the received signal for PT after sampling.
		The approximation in \eqref{y_d_P} is also similar to that in AT because $2 m v_k^{(\rm{P})} T_d  \ll d_k^{(\rm{P})}$, $\alpha n T_s  \ll  f_c$ and $\alpha \left(\tilde{T}-T \right)  \ll  f_c$. $f_{d,k}^{{(\rm{P})}}$ and $f_{v,k}^{(\rm{P})}$ are defined as the normalized distance frequency and normalized Doppler frequency for PT, which are respectively represented as
		\begin{align}
			\label{fpdvk}
			\begin{aligned}
				f_{d,k}^{{(\rm{P})}}&=N T_s \alpha d^{(\rm{P})}_k /c \in[0,N/2),\\ f_{v,k}^{(\rm{P})}&=P\bar{T}f_{c}v_{k}^{(\rm{P})}/c\in[0,P/2). 
			\end{aligned}
		\end{align}		
		
		\subsection{Design and Analysis of Data Modulation}

                \begin{table*}
	\centering
	\begin{threeparttable}
        \captionsetup{font={footnotesize}}
	\caption{Resolution and range of sensing parameters (performance bound)}
	\label{table_SensingBound}
	\begin{tabular}{lllll}
	\hline
	\multirow{2}{*}{Parameters[unit]} & \multicolumn{2}{c}{Resolution}                    & \multicolumn{2}{c}{Range}  \\ \cline{2-5} 
		& \multicolumn{1}{c|}{AT} & \multicolumn{1}{c|}{PT} & \multicolumn{1}{c|}{AT} & \multicolumn{1}{c}{PT} \\ \hline
		Distance[m]                       &$\frac{{c{{\tilde T}}}}{{2B{T}}}$&$\frac{{c{{\tilde T}}}}{{B{T}}}$&$(0, \frac{{c{T}{f_s}}}{{2(B - {f_s})}})=(0, \frac{{c(\tilde{T}-T)}}{2})$&$(0, \frac{{c{T}{f_s}}}{{(B - {f_s})}})=(0, {{c(\tilde{T}-T)}})$    \\ \hline
		Velocity[m/s]                    &$\frac{c}{{2P{{\bar T}}{f_c}}}$&$\frac{c}{{P{{\bar T}}{f_c}}}$&$(-\frac{c}{{4{{\bar T}}{f_c}}}, \frac{c}{{4{{\bar T}}{f_c}}})$&$(-\frac{c}{{2{{\bar T}}{f_c}}}, \frac{c}{{2{{\bar T}}{f_c}}})$    \\ \hline
		Elevation[1]                   &$\frac{\lambda }{{{L^{(\rm{Z})}_{\rm{tx}}}{L^{(\rm{Z})}_{\rm{rx}}}d_{a}}\cos \theta}$&$\frac{\lambda }{{{L^{(\rm{Z})}_{\rm{rx}}}d_{a}}\cos \theta}$&$(-\sin \theta_{\text{max}}, \sin \theta_{\text{max}})$&$(-\sin \theta_{\text{max}}, \sin \theta_{\text{max}})$    \\ \hline
		Azimuth[1]                     &$\frac{\lambda }{{{L^{(\rm{Z})}_{\rm{tx}}}{L^{(\rm{Z})}_{\rm{rx}}}d_{a}}\cos \varphi}$&$\frac{\lambda }{{{L^{(\rm{Z})}_{\rm{rx}}}d_{a}}\cos \varphi}$&$(-\sin \varphi_{\text{max}}, \sin \varphi_{\text{max}})$&$(-\sin \varphi_{\text{max}}, \sin \varphi_{\text{max}})$    \\ \hline
	\end{tabular}
    \end{threeparttable}
    \vspace{-3mm}
    \end{table*}

        We utilize the differential phase shift keying (DPSK) for data modulation in the complex amplitude.
		As previously mentioned, due to the pseudo-random TDM-MIMO, the slow-time indices corresponding to different transmit antennas are not evenly spaced. 
		Therefore, we consider applying differential modulation to the slow-time indices corresponding to each transmit antenna. 
		Each of the $L_{\mathrm{tx}}$ antennas transmits ${P}/{L_{\mathrm{tx}}}$ symbols, and $p' = \{0,1, \ldots, {P}/{L_{\mathrm{tx}}}-1\}$ is the index. In the $m$th group, the $p'$th symbol for the $l_{\mathrm{tx}}$th transmit antenna, $\beta_{m,p',l_{\mathrm{tx}}}^{(\mathrm{D})}$, is formulated as 
		\begin{align}
			\beta_{m,p',l_{\mathrm{tx}}}^{(\mathrm{D})} = 
			\begin{cases} 
				1, &  p' = 0 \\
				\beta_{m,p'-1,l_{\mathrm{tx}}}^{(\mathrm{D})} S_{m,p'-1,l_{\mathrm{tx}}}, & \rm{others}
			\end{cases},
		\end{align}
		where $S_{m,p'-1,l_{\mathrm{tx}}} = \prod_{i=0}^{p'-1}\beta_{m,i-1,l_{\mathrm{tx}}}^{(\mathrm{D})}$.
		Because of the utilization of the same transmit and receive antennas, there is no need for spatial phase difference pre-compensation.
		
		In Table {\rm{I}}, we present the resolution and range of the sensing parameters. Notably, compared with the traditional TDM-MIMO, the maximum unambiguous velocity has increased by a factor of $L_{\rm{tx}}$.
        Additionally, some basic value ranges are provided for $f_{d,k}^{{({\cdot})}}$ and $f_{d,m,p}^{(\mathrm{D})}$ in \eqref{fadvk}, \eqref{fpdvk}, and \eqref{fdmp}. 
		Due to the Chirp-DMA, the data modulation in the equivalent distance domain must be constrained, i.e., $f_{d,k}^{{(\cdot)}}+f_{d,m,p}^{(\mathrm{D})}<N$.
		Given the limitation, we set $f_{d,k}^{{({\rm{A}})}} \in [0, {N}/{2})$ and $f_{d,m,p}^{(\mathrm{D})} \in [0, {N}/{2})$ for stability of both sensing and communication. For demodulation, it is necessary to ensure that the data modulation interval exceeds the maximum variation in parameters between two consecutive estimations.
        Specifically, the velocity may cause a slight change in the distance between different chirps.
        Therefore, a constraint should be imposed on the equivalent distance domain modulation interval $\Delta{f}^{(\mathrm{D})}$, which is expressed as
        \begin{align}
            \Delta{f}^{(\mathrm{D})} >\alpha N T_{s} \frac{2 v^{(\mathrm{A})}_{\max} P \bar{T}}{c},
        \end{align}
        where $v^{(\mathrm{A})}_{\max}$ is the maximum velocity for AT, which can refer to the parameter estimation boundaries in Table {\rm{I}}.
        Additionally, complex amplitude modulation interval $\Delta{\beta}^{(\mathrm{D})}$ can be expressed as
        \begin{align}
            \Delta{\beta}^{(\mathrm{D})} >e^{j4\pi {\bar{p}}\bar{T}f_{c}\Delta{v_{k}^{(\rm{A})}}/c},
        \end{align}
        where ${\bar{p}} < 2 L_{\rm{tx}}$ is the slow-time interval between two uses of the same transmit antenna, $\Delta{v_{k}^{(\rm{A})}}$ denotes the absolute difference between the obtained velocity and true velocity.
        
		However, achieving perfect parameter estimation is often not feasible. Consequently, the data modulation interval can be appropriately relaxed to enhance the robustness of demodulation.
			In this paper, DFT and CS are employed. Thus, the equivalent distance modulation intervals are designed to align with the grids, $f_{d,m,p}^{(\mathrm{D})} \in \{0,1,\ldots, N/2-1 \}$.
		Therefore, the number of bits transmitted in one frame is represented as
		\begin{align}
			N_{\rm{bit}}=M P \left\lfloor\log_2(N/2)\right\rfloor+ M(P-L_{\rm{tx}})\left\lfloor\log_2{D}\right\rfloor ,
		\end{align}
		where $D$ is the order of DPSK. Consequently, the bit rate for the ISAC system is $N_{\rm{bit}}/(P\bar{T})$ bits per second (bps). 

		\begin{figure*}
			\captionsetup{font={footnotesize}, name = {Fig.}, labelsep = period}
			\centering
			\includegraphics[width=18cm, keepaspectratio]
			{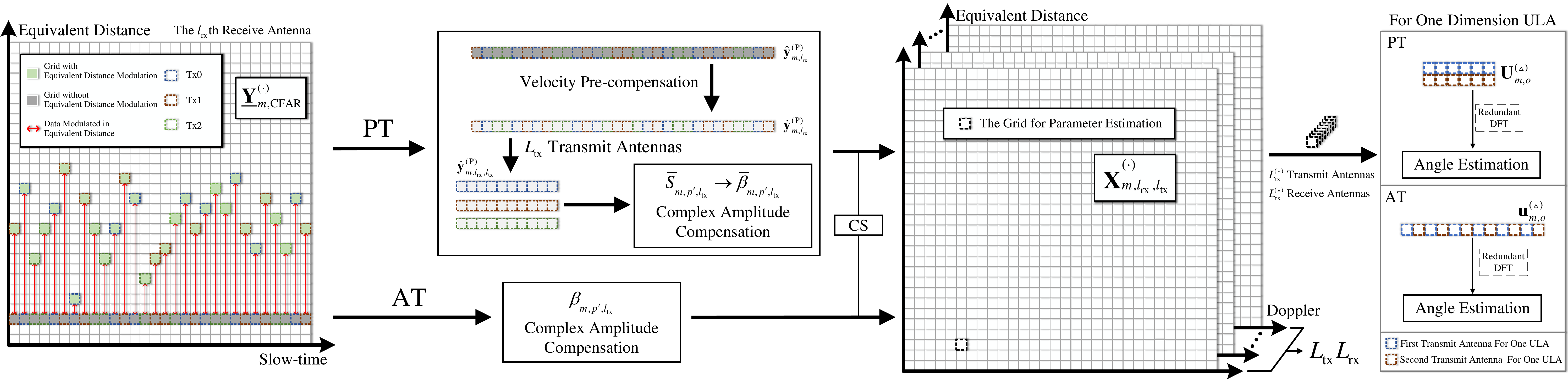}
			\caption{The schematic diagram of parameter estimation and demodulation for the proposed ISAC system assuming $M=1$ and $K^{(\rm{A})}=K^{(\rm{P})}=1$.} 
			\vspace{-6mm}
		\end{figure*}

		\section{Radar Parameter Estimation and Data Demodulation of Proposed ISAC Signal}
		Fig. 7 provides a simplified illustration of the ISAC schemes for AT and PT, assuming $M=1$, $K^{(\rm{A})}=K^{(\rm{P})}=1$, and without spectral leakage.
        For the signal processing, there are similarities and differences between AT and PT. For example, AT only requires sensing, while PT needs to perform ISAC.
        Consequently, both AT and PT initially estimate the  ``composite'' parameters, which couple radar parameters and communication data.
        Specifically, after acquiring ${\mathbf{Y}_{m,{l_{\mathrm{rx}}}}^{(\cdot)}}$ in \eqref{y_d_A} or \eqref{y_d_P}, DFT is performed to obtain the equivalent distance estimation matrix $\underline{\mathbf{Y}}_{m,l_{\mathrm{rx}}}^{(\cdot)} \in \mathbb{C}^{N \times P}$, which is represented as
		\begin{align}
			\label{range1}
	\underline{\mathbf{Y}}_{m,l_{\mathrm{rx}}}^{(\cdot)}[\hat{n},p]=\sum_{n=0}^{N-1}\mathbf{Y}_{m,l_{\mathrm{rx}}}^{(\cdot)}[n,p]e^{-j2\pi\frac{n}{N}\hat{n}},
		\end{align}
		where $\hat{n} \in \{0,1,\ldots,N-1\}$. 
The equivalent distance slow-time map ${\underline{\mathbf{Y}}}_{m}^{(\cdot)} \in\mathbb{C}^{N\times P}$ can be obtained by performing the non-coherent accumulation across $L_{\mathrm{rx}}$ receive antennas, which is represented as
		\begin{align}
{\underline{\mathbf{Y}}}_{m}^{(\cdot)}=\sum_{l_\mathrm{rx}=0}^{L_\mathrm{rx}-1}\left|{\underline{\mathbf{Y}}}_{m,l_{\mathrm{rx}}}^{(\cdot)} \right|/L_\mathrm{rx}.
		\end{align}
		Then, a one-dimensional constant false alarm rate (CFAR) detection is performed on each ${\underline{\mathbf{y}}}_{m,p}^{(\cdot)}=  {\underline{\mathbf{Y}}}_{m}^{(\cdot)}\left[{:,p}\right]\in \mathbb{C}^{N}$ to obtain the CFAR vector ${\underline{\mathbf{y}}}_{m,p,\rm{CFAR}}^{(\cdot)}\in \mathbb{C}^{N }$.
		Therefore, the CFAR matrix can be represented as ${\underline{\mathbf{Y}}}_{m,\rm{CFAR}}^{(\cdot)}= \left[{\underline{\mathbf{y}}}_{m,0,{\rm{CFAR}}}^{(\cdot)},\ldots,{\underline{\mathbf{y}}}_{m,P-1,{\rm{CFAR}}}^{(\cdot)} \right]\in \mathbb{C}^{N \times P}$.
		As shown in the leftmost grid diagram in Fig. 7,
		white grids denote $0$ and green grids denote $1$, indicating the absence or presence of a target at the grid position. 
		In each column, the green grids contain the modulated data in the equivalent distance domain and true distance estimates. 
		The red lines denote the modulated data in the equivalent distance domain, while the gray grids represent the true distance estimates. 
		
		\vspace{-4mm}
		\subsection{J-SPEDD Scheme for PT} 
		PT lacks knowledge of the modulated data, so it must first separate the predicted estimates (obtained by EKF \cite{EKF}) from ``composite'' estimates, and then perform data demodulation.
        Thereafter, PT can compensate for the data to obtain the true parameter estimates.

	\subsubsection{Equivalent Distance Demodulation}
	The ``composite'' normalized distance frequency of the $k^{({\rm{P}})}$th target in the $p$th chirp of the $m$th group is defined as ${f}_{d,m,p,k}^{({\rm{P}})}$.
	The predicted distance $\hat{d}_{k}^{(\rm{P})}$ of the $k^{({\rm{P}})}$th target is obtained from EKF and the predicted normalized distance frequency can be represented as $\hat{f}_{d,k}^{(\rm{P})} = N T_s \alpha \hat{d}_{k}^{(\rm{P})} / c$.
	Therefore, the equivalent distance-modulated data $\bar{f}_{d,m,p}^{(\mathrm{D})}$ is obtained as
	\begin{align}
		\bar{f}_{d,m,p}^{(\mathrm{D})}=\frac{1}{{{K}}^{{(\rm{P})}}}\sum_{{{k}}^{{(\rm{P})}}=1}^{{K^{{(\rm{P})}}}}\left\lfloor{f}_{d,m,p,{{k}}}^{{(\rm{P})}} - \hat{f}_{d,{{k}}}^{{(\rm{P})}}\right\rfloor,
	\end{align} 
	which refers to the red line in the leftmost grid diagram of Fig. 7.
    The extraction of distance-modulated data requires signal processing to be performed on $K^{(P)}$ targets and $P$ chirps, with a computational complexity of $\mathcal{O}(M K^{(\rm{P})} P)$.
	After equivalent distance domain compensation, the equivalent distance-modulated data removal matrix ${\mathbf{Y}}^{(\rm{P})}_{\mathrm{dc},m} \in \mathbb{C}^{N \times P}$ for PT is expressed as 
	\begin{align}
		\label{pdcom2}
		{\mathbf{Y}}^{(\rm{P})}_{\mathrm{dc},m} [\hat{n},p] = \sum_{n=0}^{N-1} e^{-j2\pi \frac{n}{N} \hat{n} {\bar{f}}_{d,m,p}^{{(\rm{D})}}}.
	\end{align} 
	Therefore, the equivalent distance slow-time map without equivalent distance-modulated data, $\overline{\mathbf{Y}}_{m,l_{\mathrm{rx}}}^{(\rm{P})} \in \mathbb{C}^{N \times P}$, can be represented as
	\begin{align}
		\label{ddcom}
		{\overline{\mathbf{Y}}}_{m,l_{\mathrm{rx}}}^{(\rm{P})} = \underline{\mathbf{Y}}_{m,l_{\mathrm{rx}}}^{(\rm{P})}\odot {\mathbf{Y}}^{(\rm{P})}_{{\mathrm{dc},m}}.
	\end{align} 
The complexity of equivalent distance compensation is $\mathcal{O}(M N L_{\rm{rx}} P)$.
Following compensation, the green grids shift in Fig. 7 downwards to gray grids, which correspond to the true distance estimates.

	
	\subsubsection{Complex Amplitude Demodulation}
	After obtaining $\overline{\mathbf{Y}}_{m,l_{\mathrm{rx}}}^{(\rm{P})}$, performing velocity pre-compensation is essential since the phase variations induced by velocity are present in each chirp, thereby affecting the demodulation of data in the complex amplitude.
	We obtain the predicted velocity by EKF, denoted as $\hat{v}_{{k}}^{(\rm{P})}$, and the predicted normalized Doppler frequency is given by $\hat{f}_{{v},{k}}^{(\rm{P})} = P\bar{T}f_{c} \hat{v}_{{k}}^{(\rm{P})}/c$.
	To streamline the processing, we isolate row vectors from  $\overline{\mathbf{Y}}_{m,l_{\mathrm{rx}}}^{(\rm{P})} $ corresponding to the $k^{(\rm{P})}$th target and consolidate them into $\widehat{\mathbf{Y}}_{m,l_{\mathrm{rx}}}^{(\rm{P})} \in \mathbb{C}^{K^{(\rm{P})} \times P}$, which is expressed as
	\begin{align}
		\label{vpc1}
		\widehat{\mathbf{Y}}_{m,l_{\mathrm{rx}}}^{(\rm{P})}\left[k^{(\rm{P})},:\right] = \overline{\mathbf{Y}}_{m,l_{\mathrm{rx}}}^{(\rm{P})} \left[\left\lfloor{f}_{d,m,p,{{k}}}^{{(\rm{P})}} - \bar{f}_{d,m,p}^{(\mathrm{D})}\right\rfloor,:\right].
	\end{align}
	By performing velocity pre-compensation, $\dot{\mathbf{Y}}_{m,l_{\mathrm{rx}}}^{(\rm{P})}  \in \mathbb{C}^{K^{(\rm{P})} \times P}$ is derived and illustrated as
	\begin{align}
		\dot{\mathbf{Y}}_{m,l_{\mathrm{rx}}}^{(\rm{P})} [{{k}}^{{(\rm{P})}},p]= \widehat{\mathbf{Y}}_{m,l_{\mathrm{rx}}}^{(\rm{P})}[{{k}}^{{(\rm{P})}},p] e^{-j2\pi \frac{p}{P}\hat{f}_{v,k}^{{(\rm{P})}}}.
	\end{align}
    The computational complexity of velocity pre-compensation is $\mathcal{O}(M L_{\rm{rx}} K^{(\rm{P})} P)$.
	For the demodulation, it is imperative to extract the column vectors corresponding to the $l_{\mathrm{tx}}$th transmit antenna and the $l_{\mathrm{rx}}$th receive antenna from all $\overline{\mathbf{Y}}_{m,l_{\mathrm{rx}}}^{(\rm{P})}$.  
    The matrix $\dot{\mathbf{Y}}_{m,l_{\mathrm{rx}},l_{\mathrm{tx}}}^{(\rm{P})} \in \mathbb{C}^{K^{(\rm{P})} \times ({P}/{L_{\mathrm{tx}}})}$ for the $l_{\mathrm{tx}}$th transmit antenna is obtained as
    \begin{align}
         {\dot{\mathbf{Y}}}_{m,l_{\mathrm{rx}},l_{\mathrm{tx}}}^{(\rm{P})} = {\dot{\mathbf{Y}}}_{m,l_{\mathrm{rx}}}^{(\rm{P})}\left[:,{{\mathbf{e}}_{m,l_{\mathrm{rx}}}}\right],   
    \end{align}
    where $\mathbf{e}_{m,l_{\mathrm{tx}}} = {\bf{E}}_{m}\left[:,l_{\mathrm{tx}}\right]$ denotes the slow-time indices corresponding to the $l_{\mathrm{tx}}$th transmit antenna for the $m$th group.

	For ${\dot{\mathbf{Y}}}_{m,l_{\mathrm{rx}},l_{\mathrm{tx}}}^{(\rm{P})}$, all ${\dot{\mathbf{Y}}}_{m,l_{\mathrm{rx}},l_{\mathrm{tx}}}^{(\rm{P})}[0,:]$ are regarded as the reference, with $\bar{S}_{m,0,l_{\mathrm{tx}}}^{(\mathrm{D})} = 1$. Thereafter, demodulation is conducted, which is represented as
	\begin{align}
		\label{ca1}
		\bar{S}_{m,p',l_{\mathrm{tx}}}^{(\mathrm{D})} = {\rm{S}} \left( \sum_{{{k}}^{{(\rm{P})}}=1}^{{K^{{(\rm{P})}}}} \sum_{l_{\mathrm{rx}}=0}^{L_{\mathrm{rx}}-1} \frac{{\dot{\mathbf{Y}}}_{m,l_{\mathrm{rx}},l_{\mathrm{tx}}}^{(\rm{P})}[k,p']}  {{\dot{\mathbf{Y}}}_{m,l_{\mathrm{rx}},l_{\mathrm{tx}}}^{(\rm{P})}[k,p'-1]}
		\right),
	\end{align}
	where ${\rm{S}} \left(\cdot\right)$ denotes the symbol decision process.
    The complexity of processing complex amplitude modulated data is $\mathcal{O}(M L_{\rm{rx}} K^{(\rm{P})} P)$.
	After demodulation, $\bar{\beta}_{m,p',l_{\mathrm{tx}}}^{(\mathrm{D})}$ can be employed for complex amplitude modulated data removal, which is expressed as 
	\begin{align}
		\label{cac1}
\bar{\beta}_{m,p',l_{\mathrm{tx}}}^{(\mathrm{D})} = 
		\begin{cases} 
			1, &  p' = 0, \\
			\bar{\beta}_{m,p'-1,l_{\mathrm{tx}}} \bar{S}_{m,p'-1,l_{\mathrm{tx}}}^{(\mathrm{D})} & \rm{otherwise}
		\end{cases}.
	\end{align}
	Therefore, ${\tilde{\mathbf{Y}}}_{m,l_{\mathrm{rx}},l_{\mathrm{tx}}}^{(\rm{P})} \in \mathbb{C}^{N \times P/L_{\rm{tx}}}$, the matrix removed both equivalent distance and complex amplitude modulated data, can be expressed as
	\begin{align}
		\label{cac2}
		\tilde{\mathbf{Y}}_{m,l_{\mathrm{rx}},l_{\mathrm{tx}}}^{(\rm{P})}[\hat{n},p'] &= \frac{\overline{\mathbf{Y}}_{m,l_{\mathrm{rx}},l_{\mathrm{tx}}}^{(\rm{P})}[\hat{n},p']}{\bar{\beta}^{(\mathrm{D})}_{m,p',l_{\mathrm{tx}}}},
	\end{align}
	and the computational complexity of this operation is $\mathcal{O}(M N L_{\rm{rx}} P)$.
    In Fig. 7, the replacement of matrix $\bf{Y}$ with vector $\bf{y}$ is because we assume ${{{K}}^{{(\rm{P})}}}=1$ in the schematic diagram. After pre-compensation, $\hat{\mathbf{y}}_{m,l_{\mathrm{rx}}}^{(\rm{P})}$ yields $\dot{\mathbf{y}}_{m,l_{\mathrm{rx}}}^{(\rm{P})}$, and it is then extracted according to the transmit antenna and undergoes demodulation and removal of the complex amplitude, and finally obtain $\tilde{\mathbf{y}}_{m,l_{\mathrm{rx}},l_{\mathrm{tx}}}^{(\rm{P})}$.
	
	\subsubsection{Parameter Estimation}
	Following the complex amplitude demodulation, we perform the true parameter estimation to obtain parameters ${d}^{(\rm{P})}_{{k}}$, ${v}^{(\rm{P})}_{{k}}$, ${{\theta}}^{(\rm{P})}_{\mathrm{tx},{{k}}}$, ${{\theta}}^{(\rm{P})}_{\mathrm{rx},{{k}}}$, ${{\phi}}^{(\rm{P})}_{\mathrm{tx},{{k}}}$, and ${{\phi}}^{(\rm{P})}_{\mathrm{rx},{{k}}}$. Initially, the CS algorithm is utilized for $\tilde{\mathbf{Y}}_{m,l_{\mathrm{rx}},l_{\mathrm{tx}}}^{(\rm{P})}$ in \eqref{cac2} to obtain distance and velocity estimates.
	The measurement matrix ${\bm{\Phi}}_{m,l_{\mathrm{tx}}}$ is constructed by selecting rows from the identity matrix ${\bf{I}}_{P} \in \mathbb{C}^{P \times P}$ corresponding to the $l_{\mathrm{tx}}$th transmit antenna, i.e., ${\bm{\Phi}}_{m,l_{\mathrm{tx}}} = {{\bf{I}}_{P} \left[{\mathbf{e}}_{m,l_{\mathrm{tx}}},: \right]} \in \mathbb{C}^{({P}/{L_{\mathrm{tx}}}) \times P}$. 
	Furthermore, a DFT matrix ${\bf{{F}}} \in \mathbb{C}^{P \times P}$ is utilized to obtain the sparse representation. Leveraging the CS framework, the estimation model can be expressed as
	\begin{align}
		\label{dv1}
		\left[{\tilde{\mathbf{Y}}}_{m,l_{\mathrm{rx}},l_{\mathrm{tx}}}^{(\rm{P})}\left[ {\hat{n}},:  \right]\right]^{\rm{T}} = {\bf{\Phi}}_{m,l_{\mathrm{tx}}} {\bf{ {F}}} \left[{{\mathbf{X}}}_{m,l_{\mathrm{rx}},l_{\mathrm{tx}}}^{(\rm{P})}\left[ {\hat{n}},: \right]\right]^{\rm{T}},
	\end{align}
	where ${{\mathbf{X}}}_{m,l_{\mathrm{rx}},l_{\mathrm{tx}}}^{(\rm{P})} \in {\mathbb{C}}^{N \times P}$ denotes the equivalent distance Doppler map for the $m$th group associated with the $l_{\mathrm{tx}}$th transmit antenna and the $l_{\mathrm{rx}}$th receive antenna. In accordance with the CS theory, if ${{\mathbf{X}}}_{m,l_{\mathrm{rx}},l_{\mathrm{tx}}}^{(\rm{P})}\left[ {\hat{n}},: \right]$ exhibits sparsity in the Doppler domain, it is feasible to reconstruct ${{\mathbf{X}}}_{m,l_{\mathrm{rx}},l_{\mathrm{tx}}}^{(\rm{P})}$ from ${\tilde{\mathbf{Y}}}_{m,l_{\mathrm{rx}},l_{\mathrm{tx}}}^{(\rm{P})}$ \cite{ke1}.
The spatial sparsity of UAVs provides a foundation for the CS reconstruction\footnote{However, in urban scenarios, the sparsity is typically low, which often leads to reconstruction failure. Under such circumstances, it is feasible to switch to the conventional TDM-MIMO interference-free multi-antenna transmission scheme illustrated in Fig. 5(c).}.
    
To date, numerous sparse signal recovery algorithms have been proposed, including $l_1$-norm minimization algorithms \cite{l1}, greedy algorithms \cite{greedy}, and iterative algorithms \cite{iterative}, among others \cite{malongke}.
In this paper, we utilize the orthogonal matching pursuit (OMP) algorithm.
OMP is a greedy algorithm used to solve the sparse reconstruction problem of the following form, which is represented as
\begin{align}
\begin{aligned}
    \min&_{{{\mathbf{X}}}_{m,l_{\mathrm{rx}},l_{\mathrm{tx}}}^{(\rm{P})}\left[ {\hat{n}},: \right]} \|{{\mathbf{X}}}_{m,l_{\mathrm{rx}},l_{\mathrm{tx}}}^{(\rm{P})}\left[ {\hat{n}},: \right]\|_0 \quad \\
    \text{subject to}& \quad \|{\tilde{\mathbf{Y}}}_{m,l_{\mathrm{rx}},l_{\mathrm{tx}}}^{(\rm{P})}\left[ {\hat{n}},:  \right] - {\bf{\Phi}}_{m,l_{\mathrm{tx}}} {\bf{ {F}}} {{\mathbf{X}}}_{m,l_{\mathrm{rx}},l_{\mathrm{tx}}}^{(\rm{P})}\left[ {\hat{n}},: \right]\|_2 \leq \epsilon,
    \end{aligned}
\end{align}
where $\epsilon$ stands for the noise tolerance.

The analysis of OMP is as follows.
\begin{itemize}
    \item In OMP, ensuring convergence requires the sensing matrix to satisfy the restricted isometry property (RIP). Since ${\bm{\Phi}}_{m,l_{\mathrm{tx}}}$ is a matrix with each row containing only one non-zero element and $\bf{F}$ is a DFT matrix, the sensing matrix satisfies the RIP with very high probability \cite{Random}.
    
    \item The computational complexity of OMP is $\mathcal{O}(M K^{\rm(P)} P \log P + M {[K^{\rm(P)}]}^2 P / L_{\mathrm{tx}})$. Though relatively higher than that of FFT schemes, CS serves as an effective trade-off strategy to increase the maximum unambiguous velocity by a factor of $L_{\rm{tx}}$ compared with the conventional TDM-MIMO scheme. Owing to the limited scale of the slow-time dimension, its computational load remains within an acceptable range.

    \item Regarding the noise robustness of the OMP, under the pseudo-random TDM-MIMO, we compare the noise robustness of velocity estimation between the FFT and OMP schemes. The FFT scheme refers to the antenna selection scheme with conventional TDM-MIMO illustrated in Fig. 5(a).  
    Specifically, we use the reconstruction success rate as the metric because parameter estimation is only related to the grid index. Furthermore, for the sake of simple comparison, the sparsity is set to $1$. The parameter settings of the OMP and FFT schemes refer to the configuration specified in the subsequent simulation section.
    As presented in Fig. 8, under low signal-to-noise ratio (SNR) conditions, the accuracy rate of OMP is lower than that of the FFT. When the SNR increases to the moderate range, OMP outperforms FFT. At high SNR, both algorithms converge to the ideal level.
    That is to say, under not particularly harsh SNR conditions, OMP exhibits better robustness against noise.
\end{itemize}

	\begin{figure}
		\captionsetup{font={footnotesize}, name = {Fig.}, labelsep = period}
		\centering
		\includegraphics[width=8cm, keepaspectratio]
		{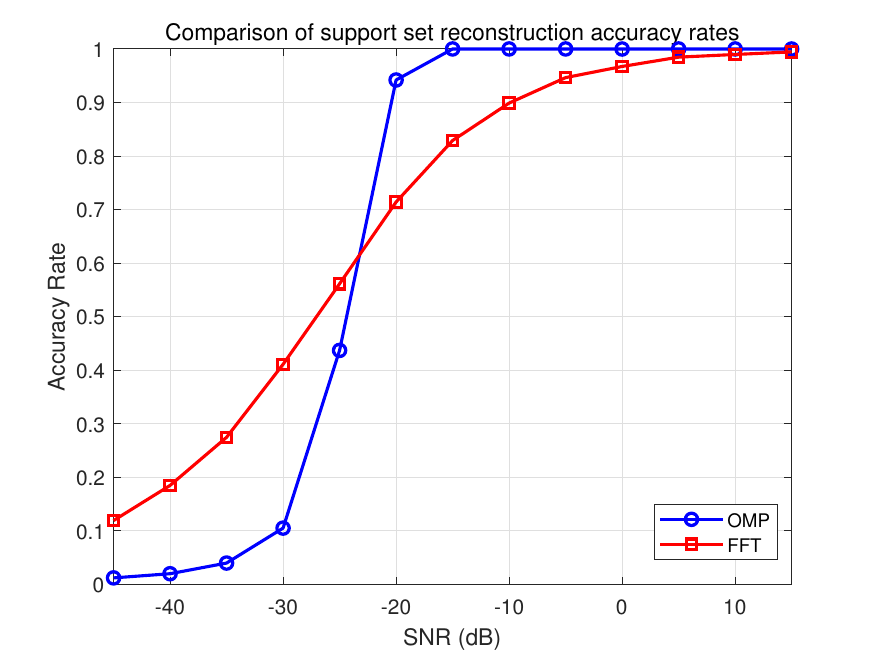}
		\caption{The noise robustness of OMP and FFT for velocity estimation.} 
		\vspace{-4mm}
	\end{figure}
    
	After CS, the non-coherent accumulation is conducted. Therefore, the equivalent distance Doppler map for the $m$th group, prior to CFAR detection,  ${\hat{\mathbf{X}}}_{m}^{(\rm{P})}\in\mathbb{C}^{N\times P}$, is defined as
	\begin{align}
		{\hat{\mathbf{X}}}_{m}^{(\rm{P})}=	\frac{1}{L}\sum_{l=0}^{L-1}\left|{\hat{\mathbf{X}}}_{m,l_{\mathrm{rx}},l_{\mathrm{tx}}}^{(\rm{P})}\right|.
	\end{align}
	Following the two-dimensional CFAR detection, we obtain the CFAR matrix ${\hat{\mathbf{X}}}_{{\rm{CFAR}},m}^{(\rm{P})} \in \mathbb{C}^{N \times P}$. 
	We perform target clustering by setting the distance and velocity thresholds $\rho_{d}$ and $\rho_{v}$.
	The elements in ${\hat{\mathbf{X}}}_{{\rm{CFAR}},m}^{(\rm{P})}$ valued as $1$ indicate the presence of targets. Supposing that there are ${O}^{(\rm{P})}$ grids as $1$, and indices are represented as $[\hat{n}^{(\rm{P})}_{{o}},\hat{p}^{(\rm{P})}_{{o}}]$, where $\hat{n}^{(\rm{P})}_{{o}}$ and $\hat{p}^{(\rm{P})}_{{o}}$ are the row and column indices of ${\hat{\mathbf{X}}}_{{\rm{CFAR}},m}^{(\rm{P})}$, respectively.
	Therefore, the criterion for target clustering is expressed as
	\begin{align}
		\label{dv2}
		\begin{aligned}
			\left| {\hat{n}^{(\rm{P})}_{{o}} - \hat{n}^{'(\rm{P})}_{{o}} } \right| \leq \rho_d \ \& \ 
			\left| { \hat{p}^{(\rm{P})}_{{o}} - \hat{p}^{'(\rm{P})}_{{o}}} \right| \leq \rho_{v},
		\end{aligned}
	\end{align}
	where $\hat{n}^{'(\rm{P})}_{{o}} \in \{0,1,\ldots,\hat{n}^{(\rm{P})}_{{o}}-1,\hat{n}^{(\rm{P})}_{{o}}+1,\ldots,{N}-1 \} $ and $\hat{p}^{'(\rm{P})}_{{o}} \in \{0,1,\ldots,\hat{p}^{(\rm{P})}_{{o}}-1,\hat{p}^{(\rm{P})}_{{o}}+1,\ldots,P-1 \} $.
	Then, equivalent distance and velocity ${{d}}^{(\rm{P})}_k$ and ${{v}}^{(\rm{P})}_k$ corresponding to the $k^{(\rm{P})}$th target can be obtained.
    
Finally, we perform the angle estimation. For angle estimation, it is necessary to extract the peaks of the equivalent distance-Doppler maps corresponding to different transmit and receive antenna pairs.
	In Section II, the relationship between $l_{\rm{tx}}^{(\triangle)}$ and $l_{\rm{tx}}$, as well as between $l_{\rm{rx}}^{(\triangle)}$ and $l_{\rm{rx}}$ are mentioned. 
    Therefore ${\hat{\mathbf{X}}}_{m,l^{(\triangle)}_{\mathrm{rx}},l^{(\triangle)}_{\mathrm{tx}}}^{(\rm{P})}$ can be extracted from ${\hat{\mathbf{X}}}_{m,l_{\mathrm{rx}},l_{\mathrm{tx}}}^{(\rm{P})}$. 
	To facilitate angle estimation, we can select antennas that form the largest ULA in one axis from all virtual antennas\footnote{This does not mean that the data from other antennas is useless. On the contrary, data from other antennas can be used to obtain diversity gain.}, as depicted in the rightmost diagram of Fig. 7. 
	The steering matrix ${\mathbf{U}}_{m,o}^{{(\triangle)}} \in \mathbb{C}^{L^{(\triangle)}_{\mathrm{tx}} \times L^{(\triangle)}_{\mathrm{rx}}}$ is expressed as
	\begin{align}
		\label{a1}
		{\mathbf{U}}_{m,o}^{{(\triangle)}}\left[l^{(\triangle)}_\mathrm{tx},l^{(\triangle)}_\mathrm{rx}\right]={\hat{\mathbf{X}}}_{m,l^{(\triangle)}_{\mathrm{rx}},l^{(\triangle)}_{\mathrm{tx}}}^{(\rm{P})} \left[\hat{n}^{(\rm{P})}_{{o}},\hat{p}^{(\rm{P})}_{{o}}\right].
	\end{align}
	Utilizing the DFT matrix with a redundant dictionary effectively reduces computational complexity while preserving high accuracy. The redundant DFT matrix is constructed for angles of departure and arrival, ${\bf{{F}}}^{{(\triangle)}}_{{\mathrm{tx}}} \in \mathbb{C}^{L^{(\triangle)}_{\mathrm{tx}} \times \mu_{{\mathrm{tx}}} L^{(\triangle)}_{\mathrm{tx}}}$ and
	${\bf{{F}}}^{{(\triangle)}}_{{\mathrm{rx}}} \in \mathbb{C}^{L^{(\triangle)}_{\mathrm{rx}} \times \mu_{{\mathrm{rx}}} L^{(\triangle)}_{\mathrm{rx}}}$, with $\mu_{{\mathrm{tx}}} \geq 1$ and $\mu_{{\mathrm{rx}}} \geq 1$, respectively, which are represented as 
	\begin{align}
		\begin{aligned}
			{\bf{{F}}}^{{(\triangle)}}_{{\mathrm{tx}}}\left[{{\hat{\chi}^{(\triangle)}_{\rm{tx}}}},{{\hat{\chi}^{(\triangle)}_{\rm{tx}, \mu}}}\right] &= e^{ {-j \pi {{\hat{\chi}^{(\triangle)}_{\rm{tx}}}} \sin \frac{({\hat{\chi}}^{(\triangle)}_{\rm{tx},\mu}-{\mu_{{\mathrm{tx}}} L^{(\triangle)}_{\mathrm{tx}}}/{2}) \pi} {\mu_{{\mathrm{tx}}} L^{(\triangle)}_{\mathrm{tx}}} } } ,\\
			{\bf{{F}}}^{{(\triangle)}}_{{\mathrm{rx}}}\left[{{\hat{\chi}^{(\triangle)}_{\rm{rx}}}},{\hat{\chi}}^{(\triangle)}_{\rm{rx}, \mu}\right] &= e^{ {-j \pi {{\hat{\chi}^{(\triangle)}_{\rm{rx}}}} \sin \frac{({\hat{\chi}}^{(\triangle)}_{\rm{tx},\mu}-{\mu_{{\mathrm{tx}}} L^{(\triangle)}_{\mathrm{rx}}}/{2}) \pi} {\mu_{{\mathrm{rx}}} L^{(\triangle)}_{\mathrm{rx}}} } } ,
		\end{aligned}
	\end{align}
	where ${{\hat{\chi}_{\rm{tx}}}} \in \{ 0,1,\ldots,L^{(\triangle)}_{\mathrm{tx}}-1 \}$, ${{\hat{\chi}_{\rm{rx}}}} \in \{ 0,1,\ldots,L^{(\triangle)}_{\mathrm{rx}}-1 \}$, ${{\hat{\chi}^{(\triangle)}_{\rm{tx},\mu}}} \in \{ 0,1,\ldots,\mu_{\rm{tx}} L^{(\triangle)}_{{\mathrm{tx}}-1} \}$, and ${{\hat{\chi}}_{\rm{rx},\mu}^{(\triangle)}} \in \{ 0,1,\ldots,\mu_{\rm{rx}} L^{(\triangle)}_{{\mathrm{rx}}-1} \}$. These two redundant matrices are used for achieving angle estimation with the unit set to degrees.
    ${\underline{\mathbf{u}}}^{(\triangle)}_{{\mathrm{tx}},m,o} \in \mathbb{C}^{\mu_{{\mathrm{tx}}} L^{(\triangle)}_{\mathrm{tx}}}$ and ${\underline{\mathbf{u}}}^{{(\triangle)}}_{{\mathrm{rx}},m,o} \in \mathbb{C}^{\mu_{{\mathrm{rx}}} L^{(\triangle)}_{\mathrm{rx}}}$ are obtained as
	\begin{align}
		\begin{aligned}
			\label{a2}
			{\underline{\mathbf{u}}}^{(\triangle)}_{{\mathrm{tx}},m,o} &=  \sum\limits_{l^{(\triangle)}_{\mathrm{rx}}=0}^{L^{(\triangle)}_{\mathrm{rx}}-1}  \left[{\bf{{F}}}^{{(\triangle)}}_{{\mathrm{tx}}} \right]^{\rm{T}}{\mathbf{U}_{m,o}^{{(\triangle)}}\left[{:,l^{(\triangle)}_{\mathrm{rx}}}\right] }, \\
			{\underline{\mathbf{u}}}^{(\triangle)}_{{\mathrm{rx}},m,o} &=  \sum\limits_{l^{(\triangle)}_{\mathrm{tx}}=0}^{L^{(\triangle)}_{\mathrm{tx}}-1}  \left[{\mathbf{U}_{m,o}^{{(\triangle)}}\left[{l^{(\triangle)}_{\mathrm{tx}},:}\right] }{\bf{{F}}}^{(\triangle)}_{{\mathrm{rx}}} \right]^{\rm{T}}.
		\end{aligned}
	\end{align}
	{The above steps are shown in the rightmost diagram of Fig. 7.}
	Then, the angles can be determined by locating the peak in ${\underline{\mathbf{u}}}^{(\triangle)}_{{\mathrm{rx}},m,o}$ and ${\underline{\mathbf{u}}}^{(\triangle)}_{{\mathrm{tx}},m,o}$.
	Therefore, angle estimates ${{\theta}}^{(\rm{P})}_{\mathrm{tx},{{k}}}$, ${{\theta}}^{(\rm{P})}_{\mathrm{tx},{{k}}}$, ${{\phi}}^{(\rm{P})}_{\mathrm{tx},{{k}}}$, and ${{\phi}}^{(\rm{P})}_{\mathrm{tx},{{k}}}$ can be obtained.

	\subsection{DCA-SPE Scheme for AT}
    AT possesses the self-transmitting and self-receiving sensing capability. Based on this capability, only compensation operations on the communication data are required.
	For AT, the equivalent distance-modulated data removal matrix ${\mathbf{Y}}^{(\rm{A})}_{\mathrm{dc},m} \in \mathbb{C}^{N \times P}$ for AT is similarly as \eqref{pdcom2}, which is defined as 
	\begin{align}
		\label{disatance com}
		{\mathbf{Y}}^{(\rm{A})}_{\mathrm{dc},m} [\hat{n},p] = \sum_{n=0}^{N-1} e^{-j2\pi \frac{n}{N} \hat{n} {f}_{d,m,p}^{{(\rm{D})}}}.
	\end{align}
	Similarly to \eqref{ddcom}, the equivalent distance slow-time map without equivalent distance-modulated data, $\overline{\mathbf{Y}}_{m,l_{\mathrm{rx}}}^{(\rm{A})} \in \mathbb{C}^{N \times P}$, is represented as
	\begin{align}
		{\overline{\mathbf{Y}}}_{m,l_{\mathrm{rx}}}^{(\rm{A})} = \underline{\mathbf{Y}}_{m,l_{\mathrm{rx}}}^{(\rm{A})}\odot {\mathbf{Y}}^{(\rm{A})}_{{\mathrm{dc},m}}.
	\end{align} 
    The complexity of equivalent distance compensation is also $\mathcal{O}(MNL_{\rm{rx}} P)$.
    The above steps for AT can also be referred to in the leftmost grid diagram of Fig. 7.
	
In complex amplitude modulated data removal, $\beta_{m,p',l_{\mathrm{tx}}}^{(\mathrm{D})}$ is known to AT.
The matrix $\overline{\mathbf{Y}}_{m,l_{\mathrm{rx}},l_{\mathrm{tx}}}^{(\rm{A})} \in \mathbb{C}^{N \times  ({P}/{L_{\mathrm{tx}}})}$ is represented as ${\overline{\mathbf{Y}}}_{m,l_{\mathrm{rx}},l_{\mathrm{tx}}}^{(\rm{A})} = {\overline{\mathbf{Y}}}_{m,l_{\mathrm{rx}}}^{(\rm{A})}\left[{:,{\mathbf{e}}_{m,l_{\mathrm{tx}}}}\right]$.
Therefore, ${\tilde{\mathbf{Y}}}_{m,l_{\mathrm{rx}},l_{\mathrm{tx}}}^{(\rm{A})} \in \mathbb{C}^{N \times (P/L_{\rm{tx}})}$, which already removed distance and complex amplitude modulated data, is expressed as
	\begin{align}
		\label{va2}
		\tilde{\mathbf{Y}}_{m,l_{\mathrm{rx}},l_{\mathrm{tx}}}^{(\rm{A})}[\hat{n},p'] = \frac{\overline{\mathbf{Y}}_{m,l_{\mathrm{rx}},l_{\mathrm{tx}}}^{(\rm{A})}[\hat{n},p']}{\beta_{m,p',l_{\mathrm{tx}}}^{(\mathrm{D})}}.
	\end{align}
The complexity of complex amplitude data removal is $\mathcal{O}(MN L_{\rm{rx}} P)$.
	At this point, all data removal for AT is completed. In the parameter estimation, the procedures for estimating distance and velocity are identical to those outlined in {\it{Parameter Estimation}} for PT, as detailed in \eqref{dv1}-\eqref{dv2}. The only modification required is substituting the $(\rm{P})$ with $(\rm{A})$. 
	
	For the angle estimation, since the arrival angle of AT is identical to the departure angle, virtual array technology \cite{vr_m} can be utilized to expand the antenna aperture. Therefore, the steering vector ${\mathbf{u}}_{m,o}^{(\triangle)} \in \mathbb{C}^{L^{(\triangle)}_{\mathrm{tx}}L^{(\triangle)}_{\mathrm{rx}}}$ is defined as
	\begin{align}
		\label{s4}
		{\mathbf{u}}_{m,o}^{{(\triangle)}}\left[l^{(\triangle)}_\mathrm{rx}+L^{(\triangle)}_\mathrm{rx}l^{(\triangle)}_\mathrm{tx}\right]={\hat{\mathbf{X}}}_{m,l^{(\triangle)}_{\mathrm{rx}},l^{(\triangle)}_{\mathrm{tx}}}^{(\rm{A})} \left[\hat{n}^{(\rm{A})}_{{o}},\hat{p}^{(\rm{A})}_{{o}}\right].
	\end{align}
	The redundant DFT matrix is denoted as ${\bf{{F}}}_{\mu}^{{(\triangle)}} \in \mathbb{C}^{L^{(\triangle)}_{\mathrm{tx}}L^{(\triangle)}_{\mathrm{rx}} \times \mu L^{(\triangle)}_{\mathrm{tx}}L^{(\triangle)}_{\mathrm{rx}}}$ with $\mu \ge 1$, which is represented as 
	\begin{align}
		{\bf{{F}}}_{\mu}^{{(\triangle)}}\left[{\hat{\chi}},{\hat{\chi}_{\mu}}\right] = e^ {-j \pi {\hat{\chi}} \sin \frac{({\hat{\chi}_{\mu}}-\mu L^{(\triangle)}_{\mathrm{tx}}L^{(\triangle)}_{\mathrm{rx}}/2) \pi} {\mu L^{(\triangle)}_{\mathrm{tx}}L^{(\triangle)}_{\mathrm{rx}}} } ,
	\end{align}
	where ${\hat{\chi}} \in \left\{ 0,1,\ldots,L^{(\triangle)}_{\mathrm{tx}} L^{(\triangle)}_{\mathrm{rx}}-1 \right\}$ and ${\hat{\chi}_{\mu}} \in \left\{ 0,1,\ldots,{\mu} L^{(\triangle)}_{\mathrm{tx}} L^{(\triangle)}_{\mathrm{rx}}-1 \right\}$. This redundant matrix is also used for achieving angle estimation with the unit set to degrees.
	The above angle estimation steps for AT are also illustrated in Fig. 7.
	${\mathbf{u}}_{m,o}^{{(\triangle)}}$ after DFT, denoted as ${\underline{\mathbf{u}}}_{m,o}^{{(\triangle)}} \in \mathbb{C}^{{\mu} L^{(\triangle)}_{\mathrm{tx}} L^{(\triangle)}_{\mathrm{rx}}}$, can be represented as 
	\begin{align}
		\label{s42}
		{\underline{\mathbf{u}}}_{m,o}^{{(\triangle)}} = \left[{ {{{{\bf{ F}}_{\mu}^{{(\triangle)}}}}} }\right]^{\rm{T}}{\mathbf{u}}_{m,o}^{{(\triangle)}}.
	\end{align}
	Then, the angle can be determined by finding the peak in ${\underline{\mathbf{u}}}_{m,o}^{{(\triangle)}}$. Finally, angle estimates ${\bm{\theta}}^{(\rm{A})}_{\mathrm{tx},{{k}}}$, ${\bm{\theta}}^{(\rm{A})}_{\mathrm{tx},{{k}}}$, ${\bm{\phi}}^{(\rm{A})}_{\mathrm{tx},{{k}}}$, and ${\bm{\phi}}^{(\rm{A})}_{\mathrm{tx},{{k}}}$ can be obtained.
    Additionally, for AT, ${{\theta}}^{(\rm{A})}_{\mathrm{tx},{{k}}} = {{\theta}}^{(\rm{A})}_{\mathrm{rx},{{k}}}$ and ${{\phi}}^{(\rm{A})}_{\mathrm{tx},{{k}}} =  {{\phi}}^{(\rm{A})}_{\mathrm{rx},{{k}}}$. 
	
	\section{Simulation Results}
	To verify the effectiveness of the ISAC waveform we proposed, the key ISAC system parameter settings are detailed and enumerated below. In simulations, a swarm of $3$ UAVs is employed, and UAVs are respectively denoted as AT, PT $1$, and PT $2$.
	In the configuration of antennas, $L^{(\rm{X})}_{\rm{tx}}=L^{(\rm{Z})}_{\rm{tx}}=2$ and $L^{(\rm{X})}_{\rm{rx}}=L^{(\rm{Z})}_{\rm{rx}}=8$, therefore $L_{\rm{tx}}=3$, $L_{\rm{rx}}=15$, and $L=45$. The antenna element spacing, denoted as $d_{a}$, is configured to $0.5774 \lambda$ for precise sensing. Consequently, the field of view (FOV) for all UAVs is delimited to $[-60^{\circ}, 60^{\circ}]$.
	For the ISAC system parameter settings, the carrier frequency $f_c$ is $77$\,GHz, wavelength $\lambda$ is $0.0039$\,m, bandwidth $B$ is $640$\,MHz, and the sampling frequency $f_s$ is $20$\,MHz.
	$T= 51.2 \mu s$, which means each chirp contains $N = 1024$ sampling points.
	In light of the non-interfering multiple antenna transmission scheme with pseudo-random TDM-MIMO, $T_d=\frac{2}{60} \times 51.2 \mu s$, $\tilde{T} = \frac{64}{60} \times 51.2 \mu s$, and $\bar{T} = \frac{76}{60} \times 51.2 \mu s$. 
	In terms of frame parameter configuration, we set $P = 120$. 
	For target clustering, the thresholds are defined as $\rho_{d} = \rho_{v} = 1$.

    Two clutter models are adopted in this paper to accommodate different scenarios. In particular, for scenarios where clutter intensity is relatively low, clutter and system thermal noise are unified into a zero-mean Gaussian white noise model. This modeling strategy is justified by the central limit theorem \cite{Guass}. Clutter, by nature, arises from the superposition of a large number of random echo signals, so its statistical distribution inherently tends toward a Gaussian profile. When further superimposed with device thermal noise, this distribution still approximates a Gaussian distribution.
    
    However, such a clutter model based on the Gaussian assumption fails to accurately characterize urban environments. This is because strong urban clutter exhibits pronounced non-Gaussian characteristics, primarily due to severe local energy fluctuations caused by discrete strong scatterers (e.g., buildings) \cite{clutter1,clutter2}. Therefore, regarding urban airspace clutter, this paper considers the following characteristics: the received signal is the superposition of target echoes, clutter, and background Gaussian noise; the amplitude of clutter is significantly higher than that of target echoes; clutter in the distance domain exhibits strong correlation between adjacent distance grids and is typically distributed in short-distance regions; ground clutter sources typically exhibit zero velocity.
	Based on these characteristics, an urban clutter model is established.
	For this model, parameters are set as follows. The statistical distribution of clutter amplitude adopts the Rayleigh Distribution, with the noise variance as $1$. The average clutter amplitude is $10$ dB higher than that of target echoes. Clutter covers $40-50$ consecutive distance grids, and the distance center is randomly selected within the $20-30$ m, and clutter velocity is set to $0$ m/s.
	Additionally, since clutter obscures the desired target signal, an additional clutter filtering step is incorporated. Specifically, after performing the distance FFT, consecutive distance grids are first identified, followed by filtering processing based on the clutter grid information \cite{clutter_amp}.
    
    For benchmarks, we select the mmWave-LoRadar \cite{taoyi} and the conventional TDM-MIMO FMCW \cite{chirp}, where TDM-MIMO FMCW is the basis of our ISAC waveform and mmWave-LoRadar is an ISAC waveform that incorporates the modulation concept of LoRa \cite{LoRa}.  
	In terms of parameter settings, both TDM-MIMO FMCW and mmWave-LoRadar adopt the same parameter settings as above.
	For mmWave-LoRadar, a spreading factor $N_{\rm SF}=15$ is used to define $H = 2^{N_{\rm{SF}}} = 32768$ possible frequency offset states, thereby embedding $N_{\rm SF}=15$ bits of data in each symbol.
        \begin{figure}[htbp]
		\captionsetup{font={footnotesize}, name = {Fig.}, labelsep = period}
		\centering
		\subfigure[]
		{\includegraphics[width=8.8cm, keepaspectratio]
		{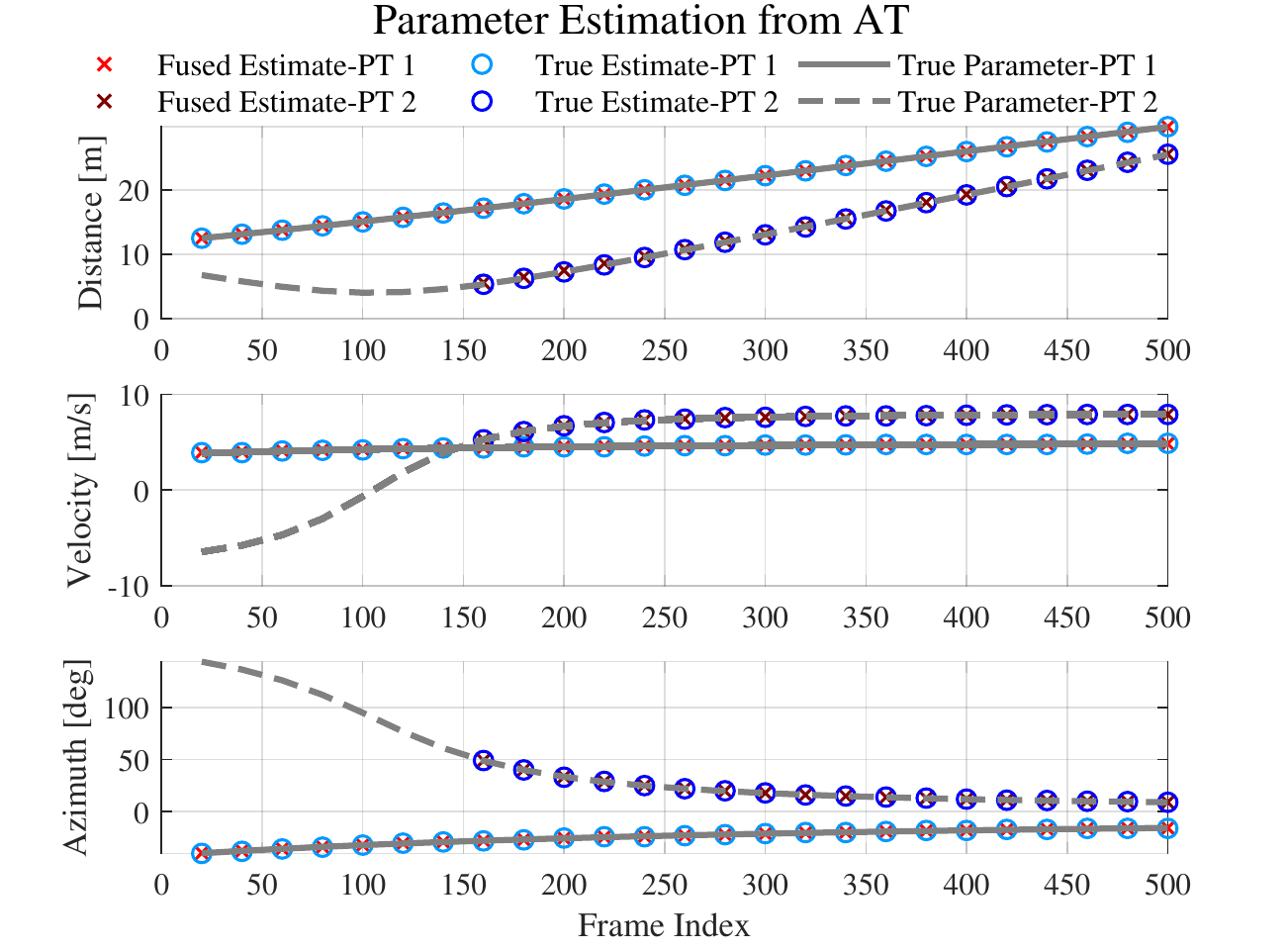}}
		\centering
		\subfigure[]{\includegraphics[width=8.8cm, keepaspectratio]
		{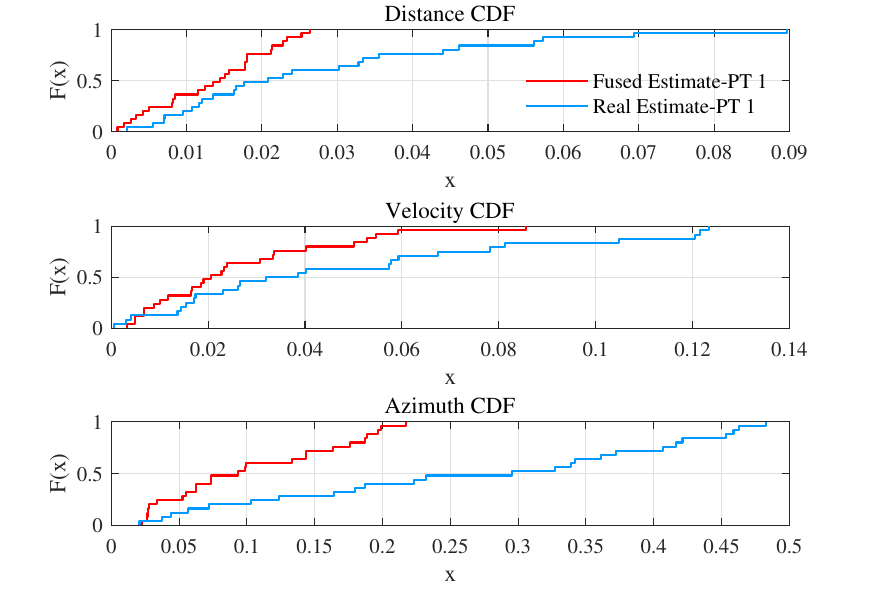}}
		\caption{The parameter estimation from AT and PT $1$ respectively in $500$ frames when SNR is fixed at $10$ dB: (a) AT; (b) PT $1$.}
		\vspace{-4mm}
	\end{figure}

	\begin{figure}[htbp]
		\captionsetup{font={footnotesize}, name = {Fig.}, labelsep = period}
		\centering
		\subfigure[]
		{\includegraphics[width=8.8cm, keepaspectratio]{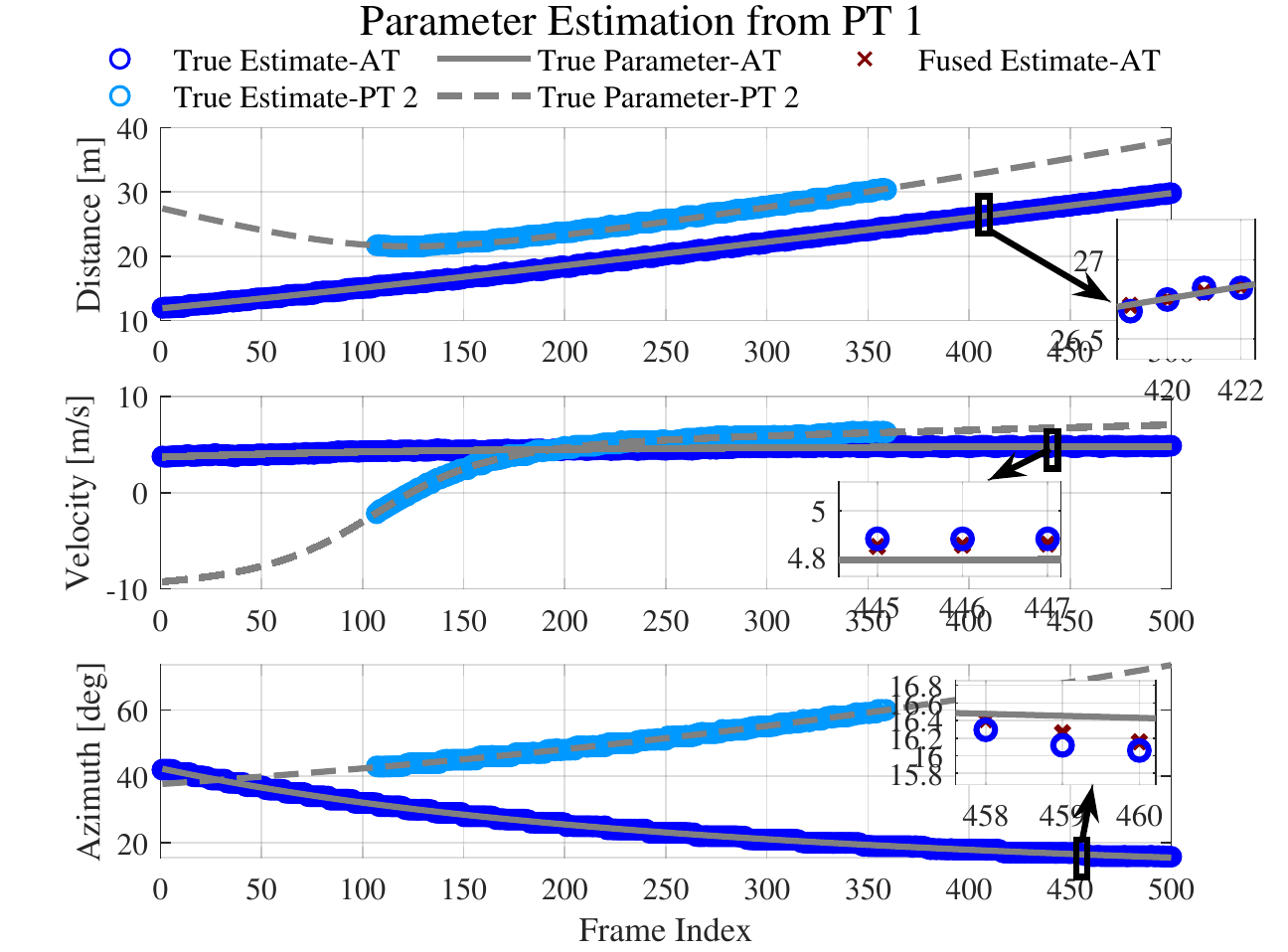}}
		\centering
		\subfigure[]
		{\includegraphics[width=8.8cm, keepaspectratio]{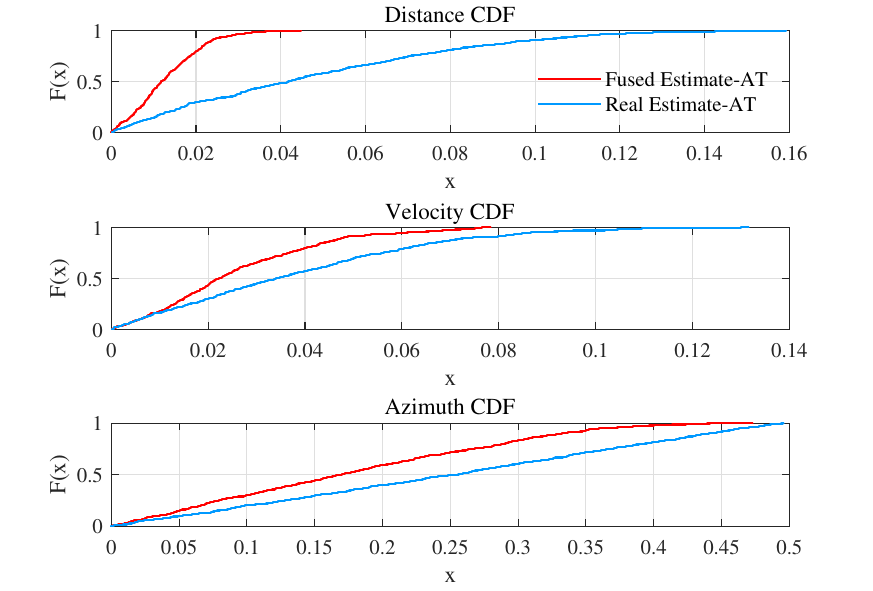}}
		\caption{The parameter estimation from AT and PT $1$ respectively in $500$ frames when SNR is fixed at $10$ dB: (a) AT; (b) PT $1$.}
		\vspace{-2mm}
	\end{figure}
	
	\begin{table}[h]
		\small
		\centering
		\caption{Initial movement settings of UAVs.}
		\begin{tabular}{cccc} %
			\toprule %
			Parameters & AT & PT $1$ & PT $2$ \\ \midrule 
			coordinates & $(0, 0, 300)$ & $(-8, 8, 300)$ & $(4, -8, 300)$\\
			velocity & $10$ m/s & $15$ m/s & $20$ m/s\\
			direction & positive $y$-axis & positive $y$-axis & negative $y$-axis\\ \bottomrule 
		\end{tabular}
        \vspace{-2mm}
	\end{table}

    In the simulations, we first verify the actual tracking performance of UAVs performing ISAC operations in a dynamic environment (see Figs. 7 and 8). Subsequently, in joint simulations, we analyze the communication performance and sensing performance separately, examine the impact of clutter, and conduct comparisons between the proposed scheme and benchmarks (see Figs. 11 to 14).

    \vspace{-3mm}
	\subsection{Dynamic ISAC in UAV Swarm}
	We first present the dynamic ISAC process from AT and PT $1$ in the UAV swarm with $3$ UAVs, demonstrating the viability of the proposed ISAC system in dynamic environments. The initial movement settings of the UAV swarm are shown in TABLE $\rm{II}$. In the ISAC process, the number of frames is set to $500$, and the SNR is fixed at $10$ dB, and does not consider the urban clutter. The tracking employs the EKF in \cite{EKF}, and the tracking intervals between frames for AT and PT $1$ are set to $20$ and $1$, respectively. This is because the requirement of AT is solely for parameter estimation optimization, whereas the PT relies on tracking for data demodulation and parameter estimation.
	
	Fig. 9(a) illustrates the dynamic parameter estimation and tracking from AT to PT $1$ and PT $2$. In this figure, ``True Parameter'' represents the true parameters of distance, velocity, and azimuth, respectively. ``True Estimate'' indicates the radar parameter estimates obtained by DFT or CS. ``Fused Estimate'' denotes the radar parameter estimates obtained from EKF.
	To demonstrate the robustness of tracking, the FOV center of AT and PT $1$ is aligned with the positive direction of the $y$-axis, whereas that of PT $2$ is oriented towards the negative direction of the $y$-axis.
	As the altitudes of UAVs remain constant, we focus on presenting the variations in the azimuth angles.
	As depicted, AT is capable of achieving precise tracking of multiple PTs. PT $1$ remains consistently within the FOV of AT. PT $2$ first enters the FOV of AT approximately at the $160$th frame, without interfering with the AT's tracking of PT $1$, which demonstrates the ISAC scheme's exceptional multi-target processing capabilities.
	In Fig. 9(b), we additionally include the cumulative distribution function (CDF) for errors between different parameters for AT, specifically encompassing the absolute errors between ``Fused Estimate'' and ``True Parameter'', as well as the absolute errors between ``True Estimate'' and ``True Parameter''.
	It can be observed that, compared with direct parameter estimation, the EKF achieves an overall leftward shift in the CDF curves through its recursive mechanism of state prediction and observation update, leading to a significant improvement in the consistency of parameter estimation.
	
	Similarly, Fig. 10(a) depicts the parameter estimation process from PT $1$ to AT and PT $2$, corresponding to the identical ISAC process shown in Fig. 7, only with the distinction in different terminals. 
For PT $1$, the impact of multipath is taken into account.
    Signals transmitted from AT may include two paths: ``AT -> PT $1$'' and ``AT -> PT $2$ -> PT $1$''. Consequently, PT $1$ is capable of obtaining parameter estimates of AT and PT $2$, as well as demodulating data transmitted by AT.
	Considering that the velocity estimation of PT $2$ relative to PT $1$ is determined by the movement of two terminals, it becomes more complex in the tracking process. Therefore, PT $1$ does not track PT $2$.
	The radar parameter estimates show no significant deviation from the true parameters, indicating that data demodulation is successfully achieved.
	Similarly, AT remains consistently within the FOV of PT $1$. PT $2$ appears between $107$th and $359$th frames and does not interfere with PT $1$'s tracking of AT. 
	PT $2$ disappears after the $359$th frame because PT $1$ is solely responsible for receiving signals and can only acquire the parameters of PT $2$ when PT $2$ is simultaneously within the FOVs of both AT and PT $1$.  
	Fig. 10(b) similarly presents the CDF curves for PT $1$, demonstrating the performance gain of the EKF tracking.
	\vspace{-3mm}
	\subsection{Data Demodulation Performance}
	In the data demodulation, we conducted a detailed investigation into the impact of parameter settings on the symbol error rates (SERs) for the complex amplitude and equivalent distance domain. In our simulations, we retained the initial coordinate settings of the UAV swarm from the tracking process, focusing on the transmission from AT to PT $1$. However, to eliminate the influence of velocity, we set the relative radial velocity to $0$ m/s. Additionally, it should be noted that the sensing simulation and communication simulation in this paper are joint simulations, meaning that parameter estimation and demodulation are performed simultaneously.

        \begin{figure}
		\captionsetup{font={footnotesize}, name = {Fig.}, labelsep = period}
		\centering
		\includegraphics[width=8.3cm, keepaspectratio]
		{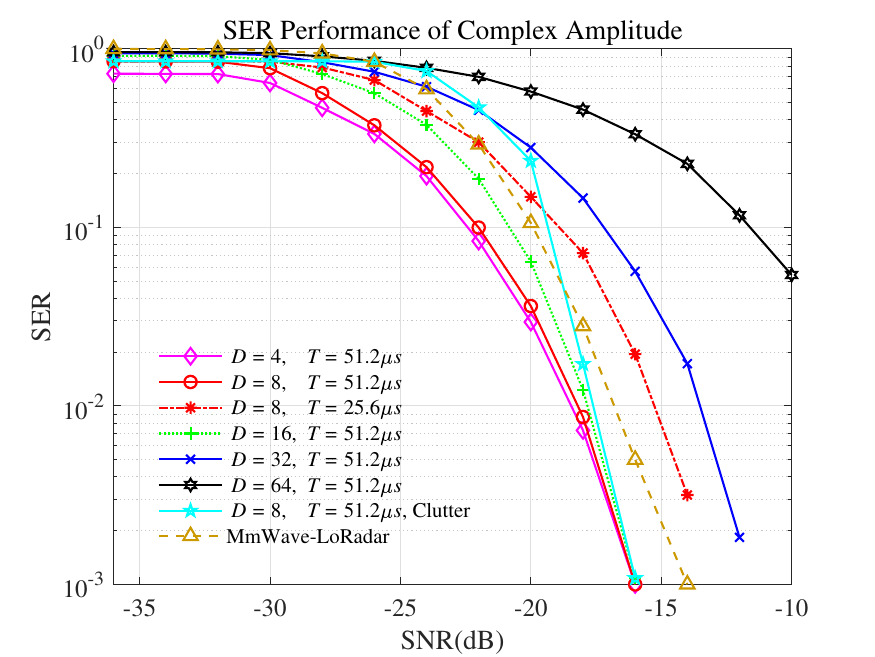}
		\caption{The SER performance of pseudo-random DPSK and the benchmark from AT to PT $1$ given $L_{\rm{tx}}=3$ and $L_{\rm{rx}}=15$.} 
		\vspace{-4mm}
	\end{figure}
	
		\begin{figure}
		\captionsetup{font={footnotesize}, name = {Fig.}, labelsep = period}
		\centering
		\includegraphics[width=8.3cm, keepaspectratio]
		{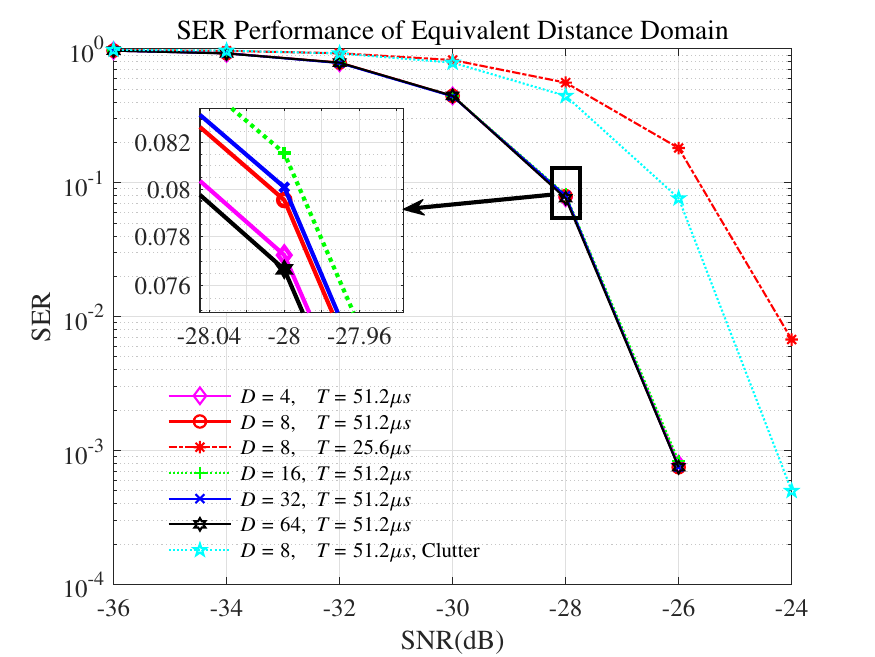}
		\caption{The SER performance of equivalent distance domain from AT to PT $1$ given $L_{\rm{tx}}=3$ and $L_{\rm{rx}}=15$.}
		\vspace{-6mm}
	\end{figure}
    
	In Fig. 11, the SER performance for the pseudo-random DPSK modulation is presented, where $D$ represents the modulation order. At extremely low SNRs, the SER plateaus exhibit variations corresponding to different $D$, as a phenomenon attributable to the distinct modulation orders.
	For the same $T=51.2 \mu s$, the SER increases as $D$ increases. 
    Additionally, we provide an SER curve for DPSK with $D=8$ and $T=25.6 \mu s$. It is apparent that a smaller $T$ leads to an increase in SER because of the decreased time-bandwidth product.
    Furthermore, the simulation results indicate that when urban spatial clutter is present, the system requires a higher SNR to achieve the same SER as that in a Gaussian noise environment. However, under high SNR conditions, the system still exhibits good robustness to the noise.
    
	Although mmWave-LoRadar employs initial frequency modulation, for fair comparison, we still present its simulation alongside the DPSK demodulation of the proposed scheme.  
	To ensure fairness, mmWave-LoRadar also uses CS for data demodulation. Specifically, when the spreading factor $N_{\rm SF} = 15$, the total number of sampling points theoretically required is $2^{N_{\rm SF}+1} = 65536$. However, by designing a sampling compression ratio of $1/64$, demodulation can be achieved with only $1024$ sampling points, which is consistent with $N$ in the proposed scheme.
	As indicated, when $D \leq 16 $, the SER performance of mmWave-LoRadar is inferior to that of the proposed scheme.
	This phenomenon is mainly attributed to two factors. First, the frequency hopping in mmWave-LoRadar causes phase discontinuities and introduces additional noise. Second, the sparsity of the mmWave channel is disturbed by multipath effects, and the communication receiver of mmWave-LoRadar does not perform real-time parameter estimation, which increases demodulation errors. In contrast, the proposed scheme significantly enhances demodulation robustness under low SNR conditions through the design of a real-time estimation and demodulation.
	Additionally, the theoretical data-carrying capacity of mmWave-LoRadar within one frame is $N_{\rm bit} = PN_{\rm SF} = 1800$ bits.
	In contrast, the proposed scheme can transmit up to $N_{\rm bit} = MP \left\lfloor\log_2(N/2)\right\rfloor + M(P-L_{\rm tx})\left\lfloor\log_2 D\right\rfloor = 4293$ bits when $ D = 8 $, which significantly outperforms the mmWave-LoRadar.

	In Fig. 12, the SER performance of equivalent distance domain modulation is presented. For $T=51.2 \mu s$, the SER remains largely unchanged with the increase of $D$, which is because the DPSK is modulated on each chirp individually, thus does not severely affect the equivalent distance domain modulation. With $D=8$ and $T=25.6 \mu s$, the time-bandwidth product is decreased, resulting in a corresponding increase in the SER.
    Similarly, urban clutter also exerts an impact on equivalent distance domain demodulation, resulting in a certain degree of degradation in its performance.

	Additionally, Fig. 13 illustrates the relationship between the transmission rate and the number of transmit antennas. With $L_{\rm{tx}}=3$ transmit antennas and the settings for $D$ and $T$ in simulations, the transmission rate ranges from $500$ to $700$ kbps.
	The transmission rate exhibits an upward trend with the increase in the number of transmit antennas, although this growth rate gradually slows down. This is because $\bar{T}$ increases with more transmit antennas. 
	\begin{figure}
		\captionsetup{font={footnotesize}, name = {Fig.}, labelsep = period}
		\centering
		\includegraphics[width=8.3cm, keepaspectratio]
		{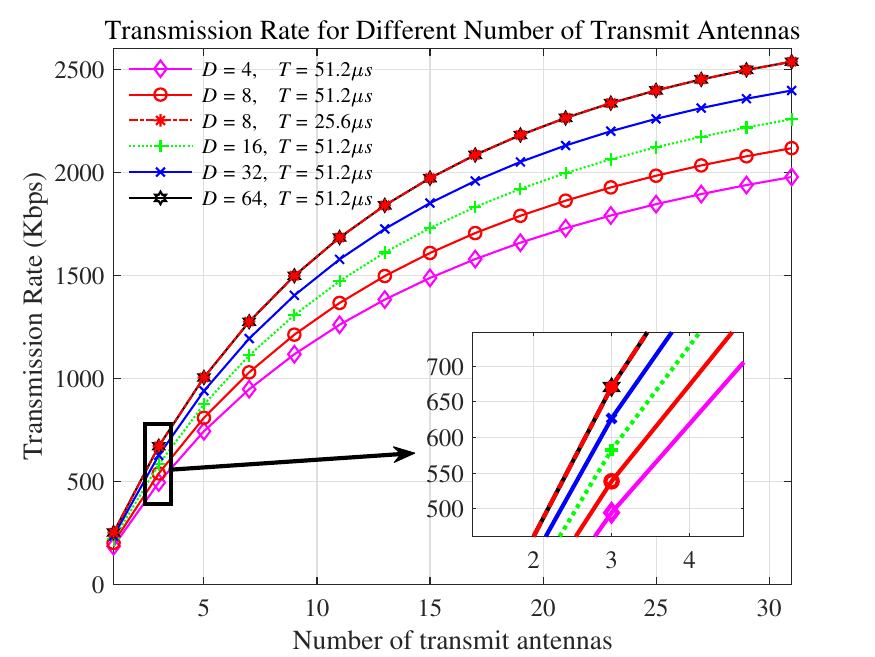}
		\caption{The relationship between the number of transmit antennas and transmission rate under the ideal noise-free environment.}
		\vspace{-5mm}
	\end{figure} 
    
	\vspace{-2mm}
	\subsection{Radar Sensing Performance}
	In sensing applications, our goal is to accurately estimate parameters within an acceptable error range.
	In terms of performance evaluation, the ``hit rate'' is considered a key metric, calculated as the ratio of successful hits to the total number of simulations.
	In this paper, a ``hit'' is defined as the simultaneous condition where errors in all parameters are less than their respective resolutions.
	Fig. 14 presents the hit rate performance under various parameter configurations.
    Simulations show that the hit rate performance is first strongly correlated with the equivalent distance demodulation performance in Fig. 12, which indicates that errors in equivalent distance demodulation will significantly affect the overall sensing performance.
	Of particular note is that the hit rate does not exhibit significant changes with variations in $D$. This stability can be attributed to the complex amplitude demodulation. After velocity pre-compensation (set to $0$m/s in the simulation, hence no compensation), the core of demodulation is actually the extraction of changes in complex amplitude. Although noise may cause errors in symbol decisions, these changes are eliminated through the complex amplitude removal processing.
	Moreover, velocity estimation is essentially an energy aggregation process in the Doppler domain, and the errors introduced by complex amplitude modulation and demodulation do not significantly affect this effect. Consequently, the increase in $D$ has a negligible impact on the hit rate performance.
    Since communication and parameter estimation are performed simultaneously, urban spatial clutter also introduces degradation on the hit rate performance.

	\begin{figure}
		\captionsetup{font={footnotesize}, name = {Fig.}, labelsep = period}
		\centering
		\includegraphics[width=8.3cm, keepaspectratio]
		{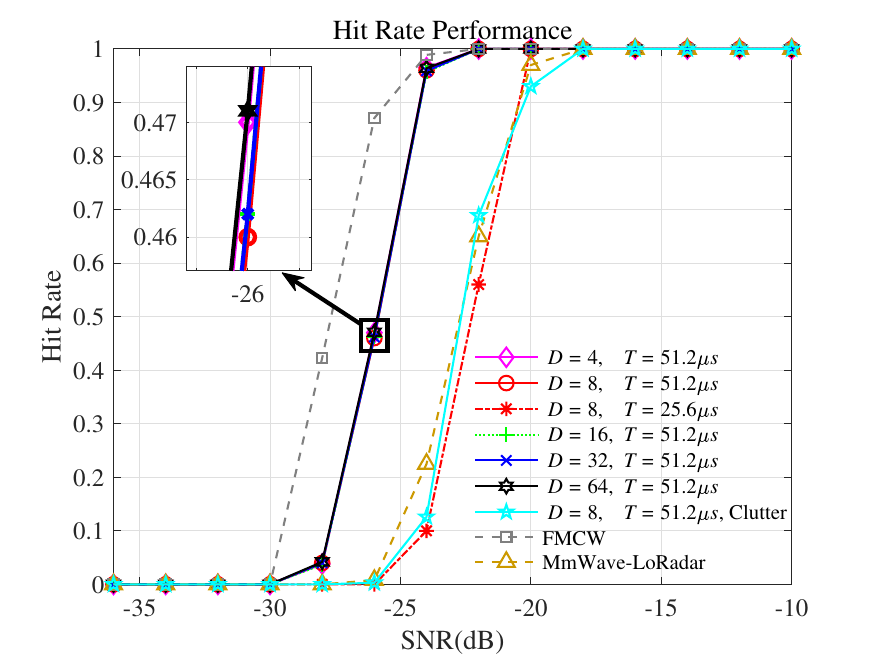}
		\caption{The hit rate performance for the proposed scheme and benchmarks when $L_{\rm{tx}}=3$ and $L_{\rm{rx}}=15$ between AT and PT $1$.}
		\vspace{-6mm}
	\end{figure}
    
	In the comparison of sensing performance, we adopted TDM-MIMO FMCW and mmWave-LoRadar as benchmarks.  
	As shown in Fig. 14, TDM-MIMO FMCW exhibits superior performance. This is because FMCW focuses exclusively on sensing functionality, and our scheme is designed based on this architecture. Furthermore, the maximum unambiguous distance of the proposed scheme is also smaller than that of FMCW.
	On the other hand, our scheme achieves a higher hit rate compared to mmWave-LoRadar. This is because in mmWave-LoRadar, symbols carrying different data have distinct initial frequencies. Thus, phase compensation is required, and error accumulation during this process leads to performance degradation. Although our proposed scheme also requires compensation, the difference in initial frequencies has a more significant impact on parameter estimation.

\section{System Performance: Challenges and Outlook}
In the simulations, we have demonstrated the feasibility of the proposed ISAC scheme and conducted basic performance analysis. However, hardware defects, high maneuverability of UAVs, and spectrum congestion will significantly affect the ISAC system performance and thus become core directions requiring focused breakthroughs in the future. 

Inherent defects of hardware components tend to cause performance degradation, which leads to the broadening of angle spectrum peaks and an increase in sidelobe levels, resulting in a reduction in detection sensitivity. The high maneuverability of UAVs also poses severe challenges to detection, as changes in UAV attitude significantly alter their radar cross section, intensifying fluctuations in target echo signals. Additionally, the problem of co-channel interference has become increasingly prominent, and such interference can be categorized into intra-chirp interference and inter-chirp interference, where Chirp-DMA can be adopted for intra-chirp interference. While for inter-chirp interference, envelope detection and differential filtering can be used.

To enhance the reliability and adaptability of the proposed ISAC system, future research should focus on developing self-compensation algorithms for hardware defects, designing intelligent tracking algorithms, and upgrading algorithms for radar cooperative sensing and dynamic resource allocation, and only through the synergistic advancement of hardware optimization, algorithm innovation, and network architecture upgrading can the multiple challenges in the future low-altitude ISAC field be effectively addressed.
    
\section{Conclusions}
	To achieve ISAC for the UAV swarm, we have proposed a pseudo-random TDM-MIMO FMCW-based ISAC solution incorporating several innovations. 
	First, we have optimized an ISAC chirp waveform that achieved efficient data modulation in terms of delay domain and complex amplitude.
	Subsequently, by combining Chirp-DMA and the pseudo-random TDM-MIMO, and utilizing different transmit antennas simultaneously, we have implemented dynamic RB allocation with multiple transmit antennas, effectively enhancing the communication rate and realizing multi-stream ISAC among multiple users, thereby expanding the system's availability.
	The pseudo-random TDM-MIMO scheme and CS algorithms have been employed to ensure the maximum unambiguous velocity and to enable the possibility of increasing the modulation order for future Doppler-based communication schemes. 
	For the proposed ISAC waveform and dynamic resource allocation scheme, we have conceived a DCA-SPE scheme for AT and a J-SPEDD scheme for PT to achieve simultaneous sensing and communication.
	Based on the above design, we have proposed a comprehensive ISAC frame structure to facilitate seamless ISAC functions.
	Simulation results indicated that the proposed waveform can achieve ISAC in the dynamic flight scenarios, fully confirming the feasibility. We also analyzed the system's joint SER and hit rate performance under Gaussian noise and urban clutter. 
    Simulation results showed that compared with the mmWave-LoRadar, the proposed scheme exhibits superior SER and hit rate performance. However, its hit rate performance is slightly lower than that of the FMCW. Furthermore, with the urban spatial clutter, although the system performance experiences a certain degree of degradation, it still demonstrates favorable robustness.
    Finally, we presented challenges and future outlook for the proposed ISAC scheme.

\end{document}